\documentclass[12pt, a4paper]{article}
\usepackage[margin = 1in]{geometry}
\usepackage[utf8]{inputenc}
\usepackage[mathscr]{eucal}
\usepackage[all]{xy}
\usepackage[dvipsnames]{xcolor}
\usepackage{authblk,amsmath,amssymb,tikz,pgfplots,arydshln,array,caption,graphicx,hyperref}
\pgfplotsset{compat=newest}
\usepgfplotslibrary{patchplots,colormaps}
\usetikzlibrary{decorations.markings}
\usepackage[section]{placeins}

\title{Representing arbitrary ground states of the toric code by a restricted Boltzmann machine}
\author[2]{Penghua Chen}
\author[2]{Bowen Yan}
\author[1,2]{Shawn X. Cui \thanks{Corresponding author}}
\affil[1]{{\small Department of Mathematics, Purdue University, West Lafayette}}
\affil[2]{{\small Department of Physics and Astronomy, Purdue University, West Lafayette}}
\affil[ ]{{\small \it \{chen3014, yan312, cui177\} @purdue.edu}}
\date{}

\begin{document}
\maketitle

\begin{abstract}
We systematically analyze the representability of toric code ground states by Restricted Boltzmann Machine with only local connections between hidden and visible neurons.  This analysis is pivotal for evaluating the model’s capability to represent diverse ground states, thus enhancing our understanding of its strengths and weaknesses. Subsequently, we modify the Restricted Boltzmann Machine to accommodate arbitrary ground states by introducing essential non-local connections efficiently. The new model is not only analytically solvable but also demonstrates efficient and accurate performance when solved using machine learning techniques. Then we generalize our the model from $Z_2$ to $Z_n$ toric code and discuss future directions.
\end{abstract}

\section{Introduction}
The research on topological phases of matter (TPMs) has significantly intensified in recent decades. These phases are characterized by topological order, setting them apart from conventional states. Topological phases feature ground states with stable degeneracy and robust long-range entanglement. In two dimensions, they support anyons and show resilience to local disruptions. These unique attributes render TPMs highly suitable for fault-tolerant quantum computing \cite{kitaev2003fault, freedman2002modular}. In two dimensions, the underlying structure of TPMs can be described by either a (2+1)-dimensional topological quantum field theory or a unitary modular tensor category. Many topological phases can be realized on spin lattice models, with the toric code model standing out as one of the most notable examples. More generally, associated with each finite group $G$, Kitaev's quantum double model defines an exactly solvable lattice model realizing possibly non-Abelian anyons. When $G=\mathbb{Z}_2$, the theory simplifies to the toric code.

Identifying the eigenstates of  the Hamiltonian of a topological phase, and more generally that of a many-body quantum system, ranks among the most demanding tasks in condensed matter physics. This task becomes increasingly complex primarily because of the power scaling of the Hilbert space dimension, which inflates exponentially in relation to the system's size \cite{osborne2012hamiltonian}. Nonetheless, it is often the case that the system's inherent physical properties, e.g. long-range entanglement, restrict the form of the ground states, and therefore the states corresponding to interesting quantum systems may only occupy a small portion of the exponentially large Hilbert space. This opens up the possibility of efficient representations of the wave function of many-body systems. Examples of efficient representations include matrix product states, projected entangled pair states, and more generally tensor networks. 

A recent trend is the study of many-body quantum systems utilizing machine learning techniques, especially artificial  neural networks. Restricted Boltzmann Machines (RBMs) are a generative stochastic artificial neural network \cite{hinton2006fast}. Unlike other types of neural networks, RBMs have a unique two-layer architecture that consists of a visible input layer and a hidden layer. The `restricted' part in the name refers to the lack of intra-layer connections; that is, nodes within the same layer do not interact with each other.  RBMs have been used effectively in a variety of machine learning tasks, including dimensionality reduction, classification, regression, and even solving quantum many-body problems \cite{carleo2017solving, deng2017quantum, deng2017machine, chen2018equivalence, gao2017efficient, lu2019efficient, jia2019efficient}.

In 2017, Carleo and Troyer paved a novel path by applying RBM as a variational ansatz, utilizing it to represent ground states for Ising model \cite{carleo2017solving}. This groundbreaking achievement catalyzed the development of numerous explicit RBM representations. Notably, substantial research efforts have been directed towards the examination of toric code \cite{deng2017quantum, deng2017machine}, tensor network state\cite{chen2018equivalence}, graph states \cite{gao2017efficient}, and stabilizer code \cite{lu2019efficient, jia2019efficient}, which is equivalent to a graph state under local Clifford operations \cite{van2004graphical}. While their topological properties and representational power \cite{le2008representational, huang2021neural} have been extensively studied, there is still a need to explore feasible algorithms for specific models.

We start from the RBM representability of the toric code model as the first step, with the eventual goal of studying that for general topological phases. In \cite{deng2017machine}, the authors utilized a Further Restricted Restricted Boltzmann Machine (FRRBM), that allows only local connections, to numerically find a ground state solution\footnote{Throughout the text, we use the term "solution" primarily to refer to the set of parameters obtained by solving the conditions that ensure the RBM represents a ground state.} of the toric code model. However, toric code has degeneracy on non-trivial topology, and the ground state derived in the above manner always corresponds certain specific one. In contrast, \cite{chen2018equivalence} presented an alternative approach, deriving a four-fold degenerate basis of ground states that can be understood by applying logical operators on the `trivial solution' \footnote{This is the simplest solution solved in Section \ref{sec:Analytical solutions of FRRBM} and further generalized in Section \ref{sec:z2 to zn}.} solved from the FRRBM. On the other hand, it is possible to achieve an arbitrary ground state by turning the toric code as a graph state \cite{liao2021graph} and transforming a graph state into an RBM \cite{lu2019efficient}. Yet, this approach inevitably introduces non-local connections within each subgraph which adds to the complexity of the RBM.

In this work, we initially apply stabilizer conditions to several specific configurations to analytically solve the FRRBM for the toric code, exploring its representational capacity. We factorize these ground state solutions on square lattices of various sizes and find that different weights only alter the coefficients of the basis states forming the ground state by factors of $\pm 1$. We then extended this approach to obtain an arbitrary ground state by strategically introducing several non-local connections into the RBM. While this generalization sacrifices the simplicity of local connections, it remains analytically solvable, enabling the simulation of arbitrary ground states in a clean manner. Additionally, we develop an efficient machine learning algorithm to verify the learnability of the models. We further generalize our approach from $Z_2$ to $Z_n$ and outline potential directions for future research.

\section{Toric code}
The toric code represents the most elementary example of Kitaev's quantum double models where $G=\mathbb{Z}_2$. It is defined on a square lattice, topologically equivalent to a torus as shown in Figure \ref{fig:2D toric lattice}. A spin $1/2$ is located on each edge $e$ of the lattice, and we also use $e$ to denote the associated qubit. The lattice components—vertices, faces, and edges—are denoted as $V$, $F$, and $E$, respectively. For each face $f \in F$, the set of surrounding edges is denoted by $s(f)$, and similarly, for each vertex $v \in V$, the set of surrounding edges is $s(v)$. The vertex operator $A_v$ consists of tensor products of Pauli operators $\hat{\sigma}_{e}^{x}$ acting on the edges $e$ within $s(v)$. Similarly, the face operator $B_f$ is formed from tensor products of Pauli operators $\hat{\sigma}_{e}^{z}$ acting on the edges $e$ within $s(f)$. The Hamiltonian of the toric code is defined as the sum of all vertex and face operators:
\begin{equation} \label{eqn:Hamiltonian of 2D toric}
    H = - \sum_{v\in V} A_v-\sum_{f \in F} B_f = - \sum_{v\in V} \prod_{e\in s(v)} \hat{\sigma}_{e}^{x} -\sum_{f \in F} \prod_{e\in s(f)}\hat{\sigma}_{e}^{z}. 
\end{equation}

\begin{table}[ht]
\centering
\includegraphics[scale=0.4]{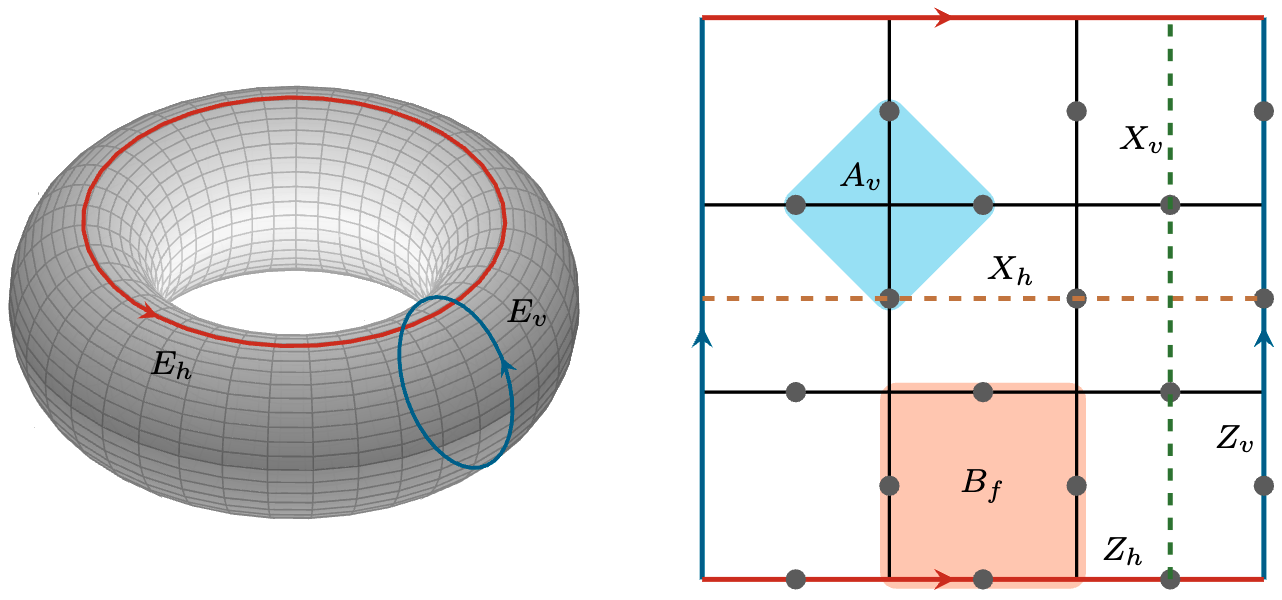}
\captionof{figure}{The torus on the left is cut along the edges $E_{v}$ and $E_{h}$ to get the square lattice shown on the right, with opposite edges identified. The $3 \times 3$ lattice shows stabilizer operators $A_{v}$ within the blue range and $B_{f}$ within the red range, logical operators $X_{v}$ and $X_{h}$ along the vertical and horizontal dashed  loops, respectively, and logical operators $Z_{v}$ and $Z_{h}$ along the edges $E_{v}$ and $E_{h}$.} 
\label{fig:2D toric lattice}
\end{table}

For a Hamiltonian of the form
\begin{equation} \label{eqn:Projector form}
    H = - \sum_{i} P_i,
\end{equation}
where each ${P_i}$ is a projector and all projectors are mutually commuting, the ground state $\lvert GS \rangle$ can be derived from any arbitrary non-zero state $\lvert \phi \rangle$: 
\begin{equation}
    \lvert GS \rangle = \prod_i P_i \lvert \phi \rangle.
\end{equation}
This construction ensures that the ground state is the simultaneous eigenvector of all projectors. Given $A_v^2 = B_f^2 = 1$ and $[A_v,B_f] = 0$ for all $v \in V$ and $f\in F$, it can be confirmed that $\frac{1+A_v}{2}$ and $\frac{1+B_f}{2}$ act as projectors. Replacing $A_v$ and $B_f$ in the Hamiltonian with these projectors yields a equivalent form consistent with Equation \ref{eqn:Projector form}. This equivalence, stemming from a one-to-one correspondence in their spectra, ensures that the state
\begin{equation}\label{eqn:Ground state of 2D toric}
    \lvert GS \rangle = \prod_{v \in V} \frac{1+A_v}{2} \prod_{f \in F} \frac{1+B_f}{2}  \lvert \phi \rangle
\end{equation}
is a valid ground state as per the earlier defined criteria. Notice that the toric code model is inherently a stabilizer code, with the local Hamiltonian terms $A_{v}$ and $B_{f}$ acting as stabilizer operators. In the context of error-correcting codes, the ground states function as logical states. The operators that dissolve these ground states are known as logical operators\footnote{$Z_h$ and $X_v$, as well as $Z_v$ and $X_h$, serve as pairs of logical $X$ and $Z$ operators for the two logical qubits, and are thus named as the logical operators.}. As described in \cite{kitaev2003fault}, the degeneracy of the ground states for a 2D toric code on a torus is identified as four distinct states: $\lvert00\rangle$, $\lvert01\rangle$, $\lvert10\rangle$, and $\lvert11\rangle$. The logical operators $X_v$ and $X_h$ consist of strings of $\hat{\sigma}_{e}^{x}$ acting along the vertical and horizontal loops, respectively, transforming $\lvert00\rangle$ to $\lvert01\rangle$ and $\lvert10\rangle$. Similarly, the logical operators $Z_v$ and $Z_h$ consist of strings of $\hat{\sigma}_{e}^{z}$ acting along $E_v$ and $E_h$, respectively, distinguish $\lvert00\rangle$ from $\lvert01\rangle$ and $\lvert10\rangle$.

\section{Restricted Boltamann Machines}

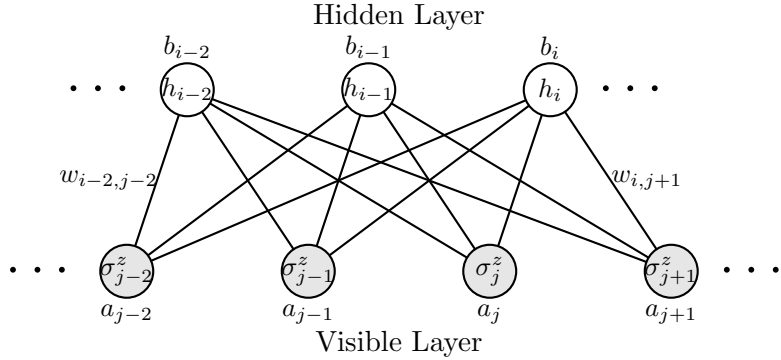
\begin{figure}[ht]
\centering
\begin{tikzpicture}[scale=0.8,
visible/.style={circle, draw=black, fill=gray!20, thick, minimum size=7mm},
hidden/.style={circle, draw=black, fill=white, thick, minimum size=7mm}]
\foreach \a in {0, 3, 6, 9}
\foreach \b in {1, 4, 7}
        \draw[thick, black] ({\a},0) -- ({\b},3);
\node at (-1-0.4,0) {\LARGE $\cdots$};
\node at (10+0.4,0) {\LARGE $\cdots$};
\node at (0-0.4,3) {\LARGE $\cdots$};
\node at (8+0.4,3) {\LARGE $\cdots$};
\foreach \i in {0, 3, 6, 9}
        \node[visible] at ({\i},0) {};
\foreach \i in {1, 4, 7}
        \node[hidden] at ({\i},3) {};
\node[black] at (1,3) {\footnotesize $h_{i-2}$};
\node[black] at (4,3) {\footnotesize $h_{i-1}$};
\node[black] at (7,3) {\footnotesize $h_{i}$};
\node[black] at (1,3.7) {\footnotesize $b_{i-2}$};
\node[black] at (4,3.7) {\footnotesize $b_{i-1}$};
\node[black] at (7,3.7) {\footnotesize $b_{i}$};
\node[black] at (0,0) {\footnotesize $\sigma_{j-2}^{z}$};
\node[black] at (3,0) {\footnotesize $\sigma_{j-1}^{z}$};
\node[black] at (6,0) {\footnotesize $\sigma_{j}^{z}$};
\node[black] at (9,0) {\footnotesize $\sigma_{j+1}^{z}$};
\node[black] at (0,-0.7) {\footnotesize $a_{j-2}$};
\node[black] at (3,-0.7) {\footnotesize $a_{j-1}$};
\node[black] at (6,-0.7) {\footnotesize $a_{j}$};
\node[black] at (9,-0.7) {\footnotesize $a_{j+1}$};
\node[black] at (-0.3,1.5) {\footnotesize $w_{i-2,j-2}$};
\node[black] at (8.6,1.5) {\footnotesize $w_{i,j+1}$};
\node[black] at (4.5,4.2) {\small Hidden Layer};
\node[black] at (4.5,-1.2) {\small Visible Layer};
\end{tikzpicture}
\caption{This diagram illustrates a RBM with visible neurons colored gray and hidden neurons colored white. The architecture ensures there are no intra-layer connections; instead, each hidden neuron is connected to all visible neurons. Each neuron and each connection is assigned a weight.}
\label{fig:RBM diagram}
\end{figure}

In the literature \cite{carleo2017solving}, Carleo and Troyer employ an Restricted Boltamann Machines (RBM) as a variational ansatz for the spin-half Ising model, as illustrated in Figure \ref{fig:RBM diagram}. The neural network consists of a layer of visible neurons corresponding to $N$ physical spins in the configuration $S =(\sigma_{1}^{z}, \sigma_{2}^{z}, \dots, \sigma_{N}^{z})$ \footnote{Throughout this paper, we use $\hat{\sigma}^{z}$ for operators and use $\sigma^{z}$ for classical variables, $\sigma^{z}=\pm 1$.}, and a single hidden layer containing $M$ auxiliary spins represented as $M = (h_{1}, h_{2}, \dots, h_{M})$. The wave function for the configuration is expressed in the variational form:
\begin{equation}
    \Psi_{M}(S;\mathcal{W})=\sum_{\{h_{i}\}}e^{\sum_{j}a_{j}\sigma_{j}^{z}+\sum_{i}b_{i}h_{i}+\sum_{ij}h_{i}w_{i,j}\sigma_{j}^{z}},
\end{equation}
where $\{ h_{i} \} = \{ -1, 1 \}^{M}$ represents all possible configurations of the hidden auxiliary spins. The network weights $\mathcal{W} = (a_{i},b_{j},w_{i,j})$ can then be trained to optimize 
\begin{equation}\label{eqn:training target}
    \lvert \Psi \rangle = \sum_{S} \Psi_{M}(S;\mathcal{W}) \lvert S \rangle
\end{equation}
to best represent the ground state $\lvert GS \rangle$. As RBM restricts intralayer interactions, we may trace out all hidden variables according to the chosen preferred basis to simplify the wave function:
\begin{equation}\label{eqn:RBM wave function}
    \Psi_{M}(S;\mathcal{W})=e^{\sum_{j}a_{j}\sigma_{j}^{z}} \prod_{i} 2\cosh(b_{i} + \sum_{j} w_{i,j} \sigma_{j}^{z}).
\end{equation}

\section{Further Restricted RBM}

\begin{table}[ht]
\centering
\begin{tabular}{m{6cm} m{5cm}}
\centering
\begin{tikzpicture}[scale=0.3]
\draw[thick, black](-4,8,-4) -- (-4,8,8) node[midway, above, sloped] {\footnotesize Visible Layer};
\draw[thick, black](4,8,-4) -- (4,8,8) node[midway, above, sloped] {\footnotesize Hidden Layer};
\node[black] at (4,4.7,-1.5) {\scriptsize $h_{v}$};
\node[black] at (4,-1.5,1.25) {\scriptsize $h_{f}$};
\node[black] at (-4,2,2.5) {\scriptsize $\sigma_{j}^{z}$};

\foreach \j in {0, 4}
        \draw[thick, black] (-4,{\j},-4) -- (-4,{\j},8);
\foreach \k in {0, 4}
        \draw[thick, black] (-4,-4,{\k}) -- (-4,8,{\k});
\foreach \j in {-4, 8}
    \draw[very thick, BrickRed](-4,{\j},-4) -- (-4,{\j},8);
\foreach \k in {-4, 8}
    \draw[very thick, MidnightBlue](-4,-4,{\k}) -- (-4,8,{\k});
\foreach \j in {-4, 0, 4} 
\foreach \k in {-2, 2, 6} 
        \shade[ball color=gray] (-4, \j, \k) circle (0.25cm);
\foreach \j in {-2, 2, 6} 
\foreach \k in {-4, 0, 4} 
        \shade[ball color=gray] (-4, \j, \k) circle (0.25cm);

\draw[very thick, dashed, Cerulean](-3.7,4+2,0) -- (4,4,0);
\draw[very thick, dashed, Cerulean](-3.7,4-2,0) -- (4,4,0);
\draw[very thick, dashed, Cerulean](-3.7,4,0+2) -- (4,4,0);
\draw[very thick, dashed, Cerulean](-3.7,4,0-2) -- (4,4,0);   
\draw[very thick, dashed, RedOrange](-3.7,-2+2,2) -- (4,-2,2);
\draw[very thick, dashed, RedOrange](-3.7,-2-2,2) -- (4,-2,2);
\draw[very thick, dashed, RedOrange](-3.7,-2,2+2) -- (4,-2,2);
\draw[very thick, dashed, RedOrange](-3.7,-2,2-2) -- (4,-2,2);

\foreach \j in {0, 4}
        \draw[thick, black] (4,{\j},-4) -- (4,{\j},8);
\foreach \k in {0, 4}
        \draw[thick, black] (4,-4,{\k}) -- (4,8,{\k});
\foreach \j in {-4, 8}
    \draw[very thick, BrickRed](4,{\j},-4) -- (4,{\j},8);
\foreach \k in {-4, 8}
    \draw[very thick, MidnightBlue](4,-4,{\k}) -- (4,8,{\k});
\foreach \i in {-4, 0, 4}
\foreach \j in {-4, 0, 4}
        \shade[ball color=Cerulean] (4, \i, \j) circle (0.25cm);
\foreach \i in {-2, 2, 6}
\foreach \j in {-2, 2, 6}
        \shade[ball color=RedOrange] (4, \i, \j) circle (0.25cm);
\end{tikzpicture}

&

\centering
\begin{tikzpicture}[scale=0.8]
\foreach \i in {0, 2}
        \draw[thick, black] ({\i},-2) -- ({\i},4);
\foreach \j in {0, 2}
        \draw[thick, black] (-2,{\j}) -- (4,{\j});
\foreach \i in {-2, 4}
    \draw[very thick, MidnightBlue, postaction={decorate, decoration={markings,
    mark=at position 0.45 with {\arrow[MidnightBlue]{stealth}}}}]
    ({\i},-2) -- ({\i},4);
\foreach \j in {-2, 4}
    \draw[very thick, BrickRed, postaction={decorate, decoration={markings,
    mark=at position 0.45 with {\arrow[BrickRed]{stealth}}}}]
    (-2,{\j}) -- (4,{\j});

\draw[very thick, dashed, Cerulean] (1,2) -- (3,2);
\draw[very thick, dashed, Cerulean] (2,1) -- (2,3);
\draw[very thick, dashed, RedOrange] (0,-1) -- (2,-1);
\draw[very thick, dashed, RedOrange] (1,-2) -- (1,0);
\foreach \i in {-1, 1, 3}
\foreach \j in {-2, 0, 2}
        \filldraw[darkgray!80] ({\i},{\j}) circle (0.1);
\foreach \i in {0, 2, 4}
\foreach \j in {-1, 1, 3}
        \filldraw[darkgray!80] ({\i},{\j}) circle (0.1);
\foreach \i in {0, 2, 4}
\foreach \j in {-2, 0, 2}
        \filldraw[Cerulean] ({\i},{\j}) circle (0.1);
\foreach \i in {-1, 1, 3}
\foreach \j in {-1, 1, 3}
        \filldraw[RedOrange] ({\i},{\j}) circle (0.1);

\node[black] at (2+0.3,1-0.3) {\scriptsize $a_{j}$};
\node[black] at (0+0.3,2-0.3) {\scriptsize $b_{v}$};
\node[black] at (1-0.3,2+0.2) {\scriptsize $w_{v,1}$};
\node[black] at (0-0.35,1+0.2) {\scriptsize $w_{v,2}$};
\node[black] at (-1+0.35,2+0.2) {\scriptsize $w_{v,3}$};
\node[black] at (0-0.35,3-0.2) {\scriptsize $w_{v,4}$};
\node[black] at (1+0.3,-1-0.3) {\scriptsize $b_{f}$};
\node[black] at (2-0.3,-1+0.2) {\scriptsize $w_{f,1}$};
\node[black] at (1-0.35,-2+0.2) {\scriptsize $w_{f,2}$};
\node[black] at (0+0.35,-1+0.2) {\scriptsize $w_{f,3}$};
\node[black] at (1-0.35,0-0.2) {\scriptsize $w_{f,4}$};
\end{tikzpicture}

\end{tabular}

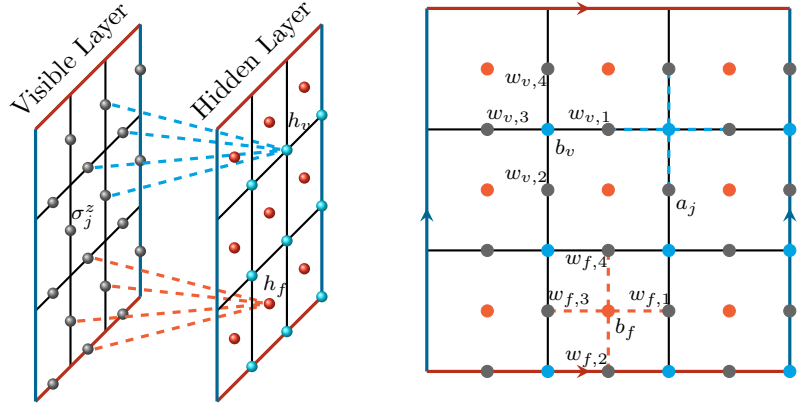
\captionof{figure}[foo]{The right diagram results from collapsing the two layers shown in the left diagram. It illustrates a translation-invariant FRRBM with visible neurons colored gray and hidden neurons colored red and blue, corresponding to faces and vertices, respectively. The architecture ensures that each hidden neuron is connected only to the nearest visible neurons. Each neuron and each connection is assigned a weight.} 
\label{tab:torus FRRBM}
\end{table}

To simulate the ground state of the 2D toric code, the authors in \cite{deng2017machine} utilized a translation-invariant Further Restricted RBM (FRRBM), illustrated in Figure \ref{tab:torus FRRBM}. This model, designed to permit only local connections, was employed to numerically find a ground state solution. For the physical spins $\{ \sigma_{j}^{z}\}$ on the square lattice, a vertex-type hidden neuron $h_v$ was assigned to each vertex and a face-type hidden neuron $h_f$ to each face, with connections limited to the nearest visible neurons. Given the two types of hidden neurons, the wave function, as shown in Equation (\ref{eqn:RBM wave function}), is reformulated as follows:
\begin{align}
    \Psi_{M}(S;\mathcal{W}) &=e^{\sum_{j}a_{j}\sigma_{j}^{z}} \prod_{v\in V} \Gamma_{v}(S;\mathcal{W}) \prod_{f\in F} \Gamma_{f}(S;\mathcal{W}), \label{eqn:FRRBM wave function}\\
    \Gamma_{v}(S;\mathcal{W})&=2\cosh(b_{v} + \sum_{j\in s(v)} w_{v,j} \sigma_{j}^{z}),\\
    \Gamma_{f}(S;\mathcal{W})&=2\cosh(b_{f} + \sum_{j\in s(f)} w_{f,j} \sigma_{j}^{z}).
\end{align}
They set weight $a_{j}=0$ for every visible neuron, choose $(b_{f}, w_{f,j})=(0,\frac{\pi}{4}i)$. Using the stabilizer conditions, they train the FRRBM to get the translational invariant and isotropic solution numerically, resulting in $(b_{v}, w_{v,j})= (0, \frac{\pi}{2}i)$. This FRRBM also naturally supports excited states if translation-invariant symmetry is broken and string operators are applied. Furthermore, this solution can be directly generalized to the 3D toric code.

\section{Analytical solutions to the FRRBM model}\label{sec:Analytical solutions of FRRBM}

However, the solution derived above is limited to describing only one specific ground state. To fully explore the representational capacity of the FRRBM, we aim to analytically solve it to identify all possible ground state solutions. We begin with the translation-invariant wave function described above and also set $a_{j}=0$ for every visible neuron. Then we solve $\mathcal{W}=(b_{f},w_{f,1-4},b_{v},w_{v,1-4})$ using the stabilizer conditions: 
\begin{align}
    B_{f} \lvert GS \rangle &= \prod_{e \in s(f)} \hat{\sigma}_{e}^{z} \lvert GS \rangle = \lvert GS \rangle, \, \forall f \in F; \label{eqn:face stabilizer condition}\\
    A_{v} \lvert GS \rangle &= \prod_{e \in s(v)} \hat{\sigma}_{e}^{x} \lvert GS \rangle = \lvert GS \rangle, \, \forall v \in V. \label{eqn:vertex stabilizer condition}
\end{align}

Solving the face stabilizer condition is straightforward, as the operator $\hat{\sigma}_{e}^{z}$ does not alter the state of the qubit $e$. By substituting Equation (\ref{eqn:FRRBM wave function}) into Equation (\ref{eqn:training target}) and treating $\lvert \Psi \rangle$ as the ground state $\lvert GS \rangle$, we solve for $\lvert \Psi \rangle$ under the constraints imposed by Equation (\ref{eqn:face stabilizer condition}). This process is applied within any single face to determine: $b_{f}=0 \, (mod \, \pi)$ and $w_{f,j}=\frac{\pi}{4}i, \frac{3\pi}{4}i \, (mod \, \pi)$, where an even number of the four $w_{f,j}$ must be the same. Further calculation details are provided in Appendix \ref{sec:appendix for face terms}.

Unlike the face stabilizer condition, solving the vertex stabilizer condition is more complex. The operator $\hat{\sigma}_{e}^{x}$ flips the state of qubit $e$. As shown in Figure \ref{tab:apply a vertex operator}, applying vertex operator to any vertex $v_{0} \in V$ alters the configuration from $\lvert S \rangle$ to $\lvert h_{0}(S) \rangle$. Notably, applying the vertex operator twice will restore the original configuration. Applying the constraints outlined in Equation (\ref{eqn:vertex stabilizer condition}), we derive the following result:
\begin{equation}
    \sum_{S} \prod_{v\in V} \Gamma_{v}(S;\mathcal{W}) \prod_{f\in F} \Gamma_{f}(S;\mathcal{W}) \lvert h_{0}(S) \rangle = \sum_{S} \prod_{v\in V} \Gamma_{v}(S;\mathcal{W}) \prod_{f\in F} \Gamma_{f}(S;\mathcal{W}) \lvert S \rangle.
\end{equation}
By applying it twice,  we can remove the sum to get
\begin{equation}\label{eqn:vertex stabilizer result}
    \prod_{v\in V} \Gamma_{v}(h_{0}(S);\mathcal{W}) \prod_{f\in F} \Gamma_{f}(h_{0}(S);\mathcal{W}) = \prod_{v\in V} \Gamma_{v}(S;\mathcal{W}) \prod_{f\in F} \Gamma_{f}(S;\mathcal{W})
\end{equation}
for any possible configuration $S$. However, there are many equal factors on both sides of the Equation (\ref{eqn:vertex stabilizer result}). Canceling out them will reduce the configuration of interest from $S$ to $S'$ which only contains 16 qubits, giving a series of $2^{16}$ equations. Directly solving these equations is impossible. We can pick up particular configurations and apply one or more vertex operators on it to get independent restrictions. Full calculation details are provided in Appendix \ref{sec:appendix for vertex terms}, solving them out, we get \footnote{We take $b_{v}=0$ as $b_{v} \in \mathbb{C}$ will introduce superfluous freedom, discussed in Appendix \ref{sec:appendix for vertex terms}.}: $b_{v}=0 \, (mod \, \pi)$ and $w_{v,j}=0, \frac{\pi}{2}i \, (mod \, \pi)$, where an even number of the four $w_{v,j}$ must be the same; Otherwise $b_{v}=0 \, (mod \, \pi)$ and any three of the four $w_{v,j}$ are equal to $0 \,\text{or}\, \frac{\pi}{2} \, (mod \, \pi)$ while the other one is free.

\begin{table}[ht]
\centering
\begin{tabular}{m{5.5cm}  m{5.5cm}} 
\centering
\begin{tikzpicture}[scale=0.6]
\foreach \i in {-4, -2, 0, 2, 4}
        \draw[thick, black] ({\i},-4) -- ({\i},4);
\foreach \j in {-4, -2, 0, 2, 4}
        \draw[thick, black] (-4,{\j}) -- (4,{\j});
\foreach \i in {-3, -1, 1, 3}
\foreach \j in {-4, -2, 0, 2}
        \filldraw[darkgray!80] ({\i},{\j}) circle (0.12);
\foreach \i in {-2, 0, 2, 4}
\foreach \j in {-3, -1, 1, 3}
        \filldraw[darkgray!80] ({\i},{\j}) circle (0.12);

\node[black] at (0+0.3,0-0.3) {\footnotesize $v_{0}$};
\end{tikzpicture}

&

\centering
\begin{tikzpicture}[scale=0.6]
\foreach \i in {-4, -2, 0, 2, 4}
        \draw[thick, black] ({\i},-4) -- ({\i},4);
\foreach \j in {-4, -2, 0, 2, 4}
        \draw[thick, black] (-4,{\j}) -- (4,{\j});
\foreach \i in {-3, -1, 1, 3}
\foreach \j in {-4, -2, 0, 2}
        \filldraw[darkgray!80] ({\i},{\j}) circle (0.12);
\foreach \i in {-2, 0, 2, 4}
\foreach \j in {-3, -1, 1, 3}
        \filldraw[darkgray!80] ({\i},{\j}) circle (0.12);

\foreach \i in {-1, 1}
\filldraw[LimeGreen] ({\i},0) circle (0.12);
\foreach \j in {-1, 1}
\filldraw[LimeGreen] (0,{\j}) circle (0.12);
\foreach \i in {-2, 0, 2}
\draw[dashed, thick, red] ({\i},0) circle (0.15);
\foreach \j in {-2, 2}
\draw[dashed, thick, red] (0,{\j}) circle (0.15);

\draw[dashed, thick, red] (0,3.5)--(3.5,0)--(0,-3.5)--(-3.5,0)--cycle;
\end{tikzpicture}

\end{tabular}

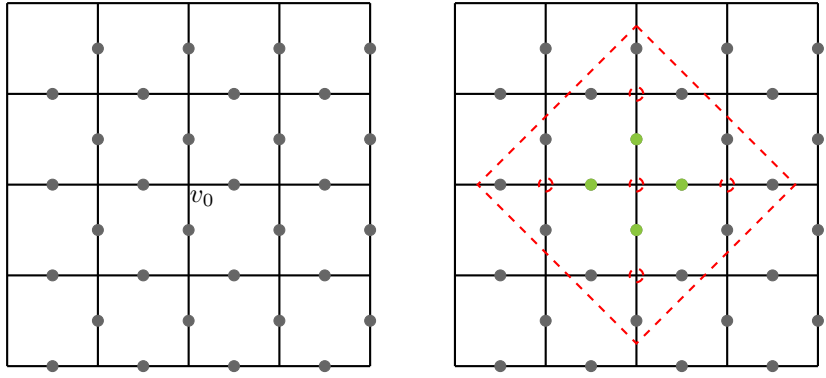
\captionof{figure}[foo]{The left diagram presents a partial view of a configuration on a larger square lattice. The right diagram is obtained by applying a vertex operator to the vertex $v_{0}$. Green nodes indicate qubits that have flipped states, and the red dashed lines encircle nodes considered in the subsequent calculation.} 
\label{tab:apply a vertex operator}
\end{table}

\section{Arbitrary ground state of RBM}

To further elucidate the analytical solutions derived in the last section, we numerically factorize them on a $3 \times 3$ square lattice, as detailed in Appendix \ref{sec:appendix for FRRBM}). Setting $(a_j,b_f,w_{f,j},b_v,w_{v,j})$ $=(0,0,\frac{\pi}{4}i,0,\frac{\pi}{2}i)$ isotropically results in the ground state $\lvert GS \rangle = -\lvert 00 \rangle + \lvert 01 \rangle + \lvert 10 \rangle -\lvert 11 \rangle$. Conversely, if we change $w_{v,j} = 0, 0, \frac{\pi}{2}i, \frac{\pi}{2}i$ for the respective directions, the ground state becomes $\lvert GS \rangle = +\lvert 00 \rangle + \lvert 01 \rangle + \lvert 10 \rangle + \lvert 11 \rangle$. Different settings of $w_{v,j}$ can alter the coefficients of the basis states forming the ground state, although these changes are confined to factors of $\pm 1$. This limitation underscores the representational capacity of the FRRBM. Consequently, it prompts a natural question: how can one prepare an arbitrary ground state?

\begin{figure}[ht]
\centering
\begin{tikzpicture}[scale=0.8]
\foreach \i in {0, 2}
        \draw[thick, black] ({\i},-2) -- ({\i},4);
\foreach \j in {0, 2}
        \draw[thick, black] (-2,{\j}) -- (4,{\j});
\foreach \i in {-2, 4}
    \draw[very thick, MidnightBlue, postaction={decorate, decoration={markings,
    mark=at position 0.45 with {\arrow[MidnightBlue]{stealth}}}}]
    ({\i},-2) -- ({\i},4);
\foreach \j in {-2, 4}
    \draw[very thick, BrickRed, postaction={decorate, decoration={markings,
    mark=at position 0.45 with {\arrow[BrickRed]{stealth}}}}]
    (-2,{\j}) -- (4,{\j});
\foreach \i in {-1, 1, 3}
        \filldraw[very thick, dashed, Orchid] (5,-3) -- (\i,-2);
\foreach \j in {-1, 1, 3}
        \filldraw[very thick, dashed, Orchid] (5,-3) -- (4,\j);
\foreach \i in {-1, 1, 3}
        \filldraw[very thick, dashed, OliveGreen] (1,-3) -- (\i,-2);
\foreach \j in {-1, 1, 3}
        \filldraw[very thick, dashed, brown] (5,1) -- (4,\j);
\foreach \i in {-1, 1, 3}
\foreach \j in {-2, 0, 2}
        \filldraw[darkgray!80] ({\i},{\j}) circle (0.1);
\foreach \i in {0, 2, 4}
\foreach \j in {-1, 1, 3}
        \filldraw[darkgray!80] ({\i},{\j}) circle (0.1);
\foreach \i in {0, 2, 4}
\foreach \j in {-2, 0, 2}
        \filldraw[Cerulean] ({\i},{\j}) circle (0.1);
\foreach \i in {-1, 1, 3}
\foreach \j in {-1, 1, 3}
        \filldraw[RedOrange] ({\i},{\j}) circle (0.1);
\filldraw[OliveGreen!70] (1,-3) circle (0.2);
\filldraw[brown!80] (5,1) circle (0.2);
\filldraw[Orchid!80] (5,-3) circle (0.2);

\node[black] at (1,-3) {\scriptsize $h_{x}$};
\node[black] at (5,1) {\scriptsize $h_{y}$};
\node[black] at (5,-3) {\scriptsize $h_{z}$};
\node[black] at (1+0.5,-3) {\scriptsize $b_{x}$};
\node[black] at (5+0.5,1) {\scriptsize $b_{y}$};
\node[black] at (5+0.5,-3) {\scriptsize $b_{z}$};
\node[black] at (-0.4,-2.6) {\scriptsize $w_{x}$};
\node[black] at (4.6,2.4) {\scriptsize $w_{y}$};
\node[black] at (4.4,-2.4) {\scriptsize $w_{z}$};
\end{tikzpicture}
\caption{Three hidden neurons ($h_x$, $h_y$, $h_z$) are introduced into the FRRBM to simulate an arbitrary ground state. $h_x$ connects to visible neurons along a horizontal loop, $h_y$ connects along a vertical loop, and $h_z$ connects to all neurons connected by $h_x$ and $h_y$. Each connection type from a specific hidden neuron is uniformly weighted ($w_x$, $w_y$, $w_z$).}
\label{tab:arbitrary torus RBM}
\end{figure}
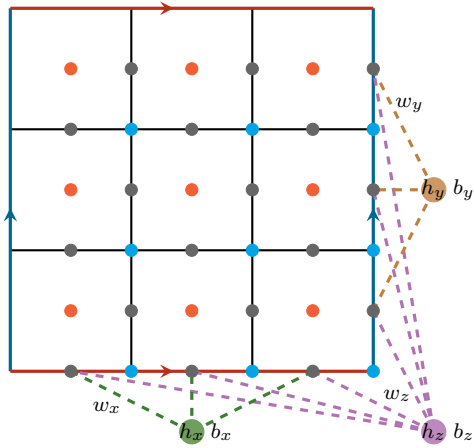

Inspired by the action of the logical operators $Z_v$ and $Z_h$,  we introduce three additional hidden neurons ($h_x$, $h_y$, and $h_z$) to the FRRBM, enabling it to encapsulate the topological information of ground states in a 2D toric code. These neurons have non-local connections as depicted in Figure \ref{tab:arbitrary torus RBM}. We demonstrate that the inclusion of $h_x$, $h_y$, and $h_z$ allows for the simulation of any arbitrary ground state. The wave function in Equation (\ref{eqn:FRRBM wave function}) is modified as follows:
\begin{align}
    \Psi_{M}(S;\mathcal{W}) \!=&\!e^{\sum_{j}\!a_{j}\sigma_{j}^{z}} \! \prod_{e\in\{x,y,z\}}\!\Gamma_{e}(S;\mathcal{W}) \prod_{v\in V} \!\Gamma_{v}(S;\mathcal{W}) \prod_{f\in F} \!\Gamma_{f}(S;\mathcal{W}), \label{eqn:the RBM wave function}\\
    &\Gamma_{e}(S;\mathcal{W})\!=\!2\cosh(b_{e}\!+\!\sum_{j}\!w_{e} \sigma_{j}^{z}).
\end{align}
If we set the parameters $(a_{j}, b_{f}, w_{f,j}, b_{v}, w_{v,j},w_{x,y,z})$ to $(0, 0, \frac{\pi}{4}i, 0, \frac{\pi}{2}i,\frac{\pi}{4}i)$,  the unnormalized ratio of the ground state on a $3 \times 3$ square lattice can be analytically derived:
\begin{align}
    \langle GS \lvert 00 \rangle &=-\cosh(b_x + \frac{\pi}{4}i)\cosh(b_y + \frac{\pi}{4}i)\cosh(b_z + \frac{\pi}{2}i), \label{eqn:ratio_00} \\
    \langle GS \lvert 01 \rangle &=\cosh(b_x - \frac{\pi}{4}i)\cosh(b_y + \frac{\pi}{4}i)\cosh(b_z), \label{eqn:ratio_01} \\
    \langle GS \lvert 10 \rangle &=\cosh(b_x + \frac{\pi}{4}i)\cosh(b_y - \frac{\pi}{4}i)\cosh(b_z), \label{eqn:ratio_10} \\
    \langle GS \lvert 11 \rangle &=-\cosh(b_x - \frac{\pi}{4}i)\cosh(b_y - \frac{\pi}{4}i)\cosh(b_z - \frac{\pi}{2}i). \label{eqn:ratio_11} 
\end{align}
For example, we can select the degeneracy state $\lvert 00 \rangle$ by setting $(b_{x}, b_{y}, b_{z})$ to $(\frac{3\pi}{4}i, \frac{3\pi}{4}i, \frac{\pi}{2}i)$. Arbitrary ground states with amplitude ratios like $\langle GS \lvert 00 \rangle\!:\!\langle GS \lvert 01 \rangle\!:\!\langle GS \lvert 10 \rangle\!:\!\langle GS \lvert 11 \rangle = 1\!:\!2\!:\!3\!:\!4$ can be exactly solved. While an exact solution for certain ratios containing $0$ may not exist, we can approximate these by setting the zeros to extremely small values. Further details are provided in Appendix \ref{sec:appendix for abitrary}. While this generalization sacrifices the simplicity of local connections, it remains analytically solvable and enables the simulation of all possible ground states in a clean manner. It also retains the ability to manipulate string operators and has demonstrated both efficient and accurate performance when applied with machine learning techniques.

\section{Efficiency and Learnability of the RBM}\label{sec:Efficiency and Learnability of the RBM}

To make $\lvert \Psi \rangle = \sum_{S} \Psi_{M}(S;\mathcal{W}) \lvert S \rangle$ best represent the ground state $\lvert GS \rangle$, we train the network weights $\mathcal{W}$ in the wave function $\Psi_M$ (as defined in Equation (\ref{eqn:the RBM wave function})) by optimizing a cost function derived from stabilizer conditions and ground state amplitude ratios. For each configuration $S$ in the selected set $\mathcal{S}$, we compute the differences in the wave function before and after applying each stabilizer operation\footnote{As $\lvert S \rangle$ is a product state, the action of $A_v$ on $\lvert S\rangle$ changes the state, whereas the action of $B_f$ only multiplies it by a number.}. The stabilizer cost function (Equation (\ref{eqn:stabilizer cost function})) is then the total sum of these differences over all selected configurations. Moreover, the cost function for an arbitrary ground state (Equation (\ref{eqn:cost function})) includes an additional term accounting for the differences between the modeled amplitudes and the expected amplitudes of the ground state.

\begin{equation}
    cost = \sum_{S\in \mathcal{S}} \left( \sum_{v}\lvert \Psi_M (S) - \Psi_M (A_v S)\rvert + \sum_{f}\lvert \Psi_M (S) - B_f\Psi_M (S)\rvert \right),\label{eqn:stabilizer cost function}
\end{equation}
\begin{equation}
    COST = cost + \lambda\sum_{S\in \mathcal{S'}}\lvert\langle \Psi \lvert S \rangle - \langle GS \lvert S \rangle \rvert,\label{eqn:cost function}
\end{equation}
where $\lambda$ is an adjustable parameter and $\mathcal{S'}$ is a subset of configurations used to characterize $\lvert GS \rangle$, as detailed Appendix \ref{sec:appendix for abitrary}.

\begin{figure}[ht]
  \centering
  \begin{minipage}[b]{0.48\textwidth}
    \centering
    \includegraphics[width=\textwidth]{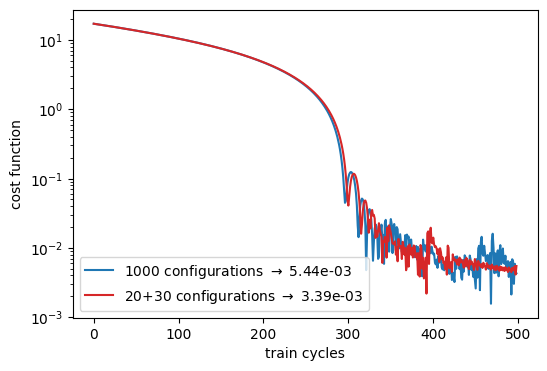}
  \end{minipage}
  \hfill
  \begin{minipage}[b]{0.48\textwidth}
    \centering
    \includegraphics[width=\textwidth]{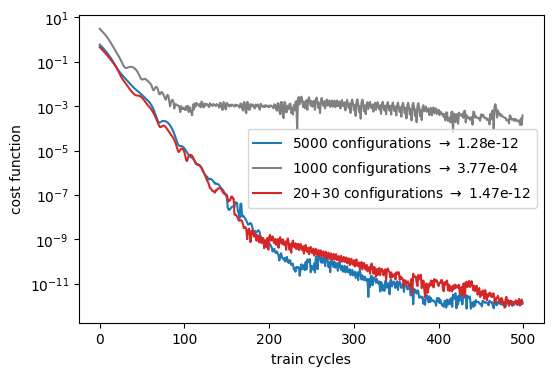}
  \end{minipage}
  \caption{On a $3\times3$ square lattice, we separately train the face terms and vertex terms using face stabilizer and vertex stabilizer conditions, respectively. The left plot compares the training efficiencies of large number of random configurations and the selected configurations for face terms, while the right plot does the same for vertex terms.}\label{fig:FRRBM different configurations comparison.}
\end{figure}

In prior work \cite{deng2017machine}, the authors analytically derived a solution for the face terms and trained the FRRBM using a vertex stabilizer condition on a portion of a larger square lattice, employing a large number of random configurations. In contrast, our study analytically derives both face and vertex terms and numerically verifies their learnability on square lattices of various sizes using a significantly reduced set of configurations $\mathcal{S}$. For example, as demonstrated in Figure \ref{fig:FRRBM different configurations comparison.}, only 50 configurations—20 selected for the degeneracy basis and 30 random configurations—are required to achieve high precision on a $3 \times 3$ square lattice. This approach is both efficient and accurate compared to methods that rely on large numbers of random configurations.

\begin{figure}[ht]
  \centering
  \begin{minipage}[b]{0.48\textwidth}
    \centering
    \includegraphics[width=\textwidth]{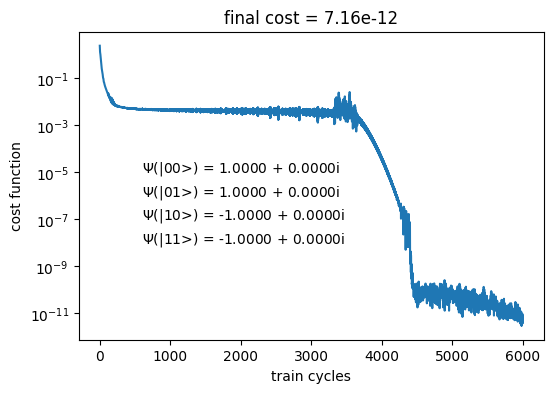}
  \end{minipage}
  \hfill
  \begin{minipage}[b]{0.48\textwidth}
    \centering
    \includegraphics[width=\textwidth]{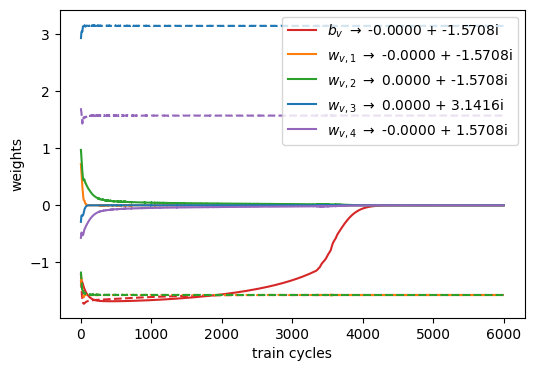}
  \end{minipage}
  \caption{On a $3\times3$ square lattice, this example demonstrates the existence of barren plateaus, characterized by a sudden drop in performance after prolonged training.}
  \label{fig:barren plateaus}
\end{figure}

As learnability is influenced by the initial settings, we randomly select 10,000 settings for $b_v \in \mathbb{C}$ and $w_v \in \mathbb{C}$ to numerically search for solutions to the constraints. Despite this extensive search, the presence of \textit{Barren Plateaus}, illustrated in Figure \ref{fig:barren plateaus}, limited us to only 20 solutions. Although the search parameters $b_v$ and $w_v$ were complex, we found solutions only where both $b_v$ and $w_v$ are purely imaginary. Barren plateaus are regions in the optimization landscape where gradients vanish, impeding any significant learning progress. This phenomenon explains why our search procedure, with limited training time, only yielded a few solutions.

\section{Generalization from \texorpdfstring{$\mathbb{Z}_2 \text{ to } \mathbb{Z}_n$}{Z2 to Zn}}\label{sec:z2 to zn}

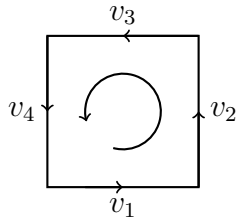
\begin{figure}[ht]
\begin{center}
\begin{tikzpicture}[scale=1.0]
\draw[black, thick] (-1,-1) rectangle (1,1);
\draw[->, black, thick] (-0.5,-1)--(0,-1) node[anchor=north] {$v_1$};
\draw[->, black, thick] (1,-1)--(1,0) node[anchor=west] {$v_2$};
\draw[->, black, thick] (1,1)--(0,1) node[anchor=south] {$v_3$};
\draw[->, black, thick] (-1,1)--(-1,0) node[anchor=east] {$v_4$};
\draw[->, black, thick] (-0.12941,-0.48296) arc (-105:195:0.5);
\end{tikzpicture}
\end{center}
\caption{Convention for the local operator: Edges are ordered counterclockwise as $v_1, v_2, v_3, v_4$, with directions indicated by arrows on each edge.}
\label{fig:definition of B}
\end{figure}

In previous sections, we determined the weights of the RBMs analytically and numerically to assess their representational capabilities. Notably, $B_f$ selects configurations with trivial flux that can survive, while $A_v$ ensures that all configurations have uniform weight across a logical state. Among the ground state solutions we found, a `trivial solution' with $(b_f, w_{f,j},b_{v}, w_{v,j})=(0,\frac{\pi}{4}i,0,0)$ emerged, where each survived configuration possess equal weight. We observe that ignoring vertex-type hidden neurons in the FRRBM still provides a good solution. When $B_f$ selects survival configurations, it assigns each configuration an equal weight, thereby automatically fulfilling the action of $A_v$. This observation receives ancillary support from the work \cite{chen2018equivalence}, where an FRRBM was constructed without using face-type hidden neurons, equivalent to our approach when aligned with our notation. This idea can be generalized to implement the \textit{Kitaev quantum double model} associated with the group $Z_N$\footnote{For $N=2$, the model corresponds to the toric code with a basis change to match notation in previous sections.}. The model, set on an oriented lattice with a $|G|$-dimensional qudit on each edge labeled by a group element $g$, follows the convention in \cite{yan2022ribbon}. Though the Hamiltonian resembles Equation (\ref{eqn:Hamiltonian of 2D toric}), $A_v$ and $B_f$ are defined in a different manner. As shown in Figure \ref{fig:definition of B}, we focus exclusively on the action of $B_f$:
\begin{equation}
    B_f \lvert v_{1} \; v_{2} \; v_{3} \; v_{4}\rangle = \delta_{p_g, 1_g} \lvert v_{1} \; v_{2} \; v_{3} \; v_{4}\rangle,
\end{equation}
where $1_g$ is the identity element of the group $G$, and $p_g$ is the group product of states on each edge bordering the face counterclockwise\footnote{We need to pick up a start-up vertex, though it turns out to be insignificant.}. If an edge’s direction aligns with the orientation, we include $v_i$; otherwise, we use $v_i^{-1}$. Thus, $p_g = \prod_i v_i^{a_i}$, where $a_i=\pm 1$ reflects this alignment. Specifically, for $G = Z_N$, the state of the $N$-dimensional qudit is labelled by $0,1,\dots N-1$.  In this setting, the group product is arithmetic summation, the identity element $1_g$ is $0$, and each element is its own inverse. Then the action of $B_f$ is significantly simplified:
\begin{equation}
    B_f \lvert v_{1} \; v_{2} \; v_{3} \; v_{4}\rangle = \delta_{\sum_{i} v_{i},0} \lvert v_{1} \; v_{2} \; v_{3} \; v_{4}\rangle.
\end{equation}

We utilize a natural RBM to implement the above action, corresponding to the condition $\sum v_i = 0 \, (mod \, N)$. This is achieved using an $N$-dimensional invisible qudit $u$ with the setting of isotropic weights $(b,w_i)=(0,\frac{2 \pi i}{N})$:
\begin{equation}
   \sum_{u=0}^{N-1} \exp(b+\sum_i u w_i v_i) = \frac{1 - \exp(2 \pi \mathrm{i} (b+\sum_i v_i))}{1 - \exp\left(2 \pi \mathrm{i} (b+\sum_i v_i) /N \right)} =
  \begin{cases} 
    N & \text{if }  b+\sum_i v_i= 0 \, (mod \, N).\\
    0 & \text{otherwise.}
  \end{cases}
\end{equation}
The action of $A_v$ is safely neglected here, as the RBM already simulates a ground state that is an equal superposition of all logical bases. And this expression explicitly ensures the flux-free requirement.  A complete basis of the ground state can be found in the same manner as illustrated in Figure \ref{tab:arbitrary torus RBM}. It offers a natural method for creating fluxions by setting $b\neq 0$, though such creation is not arbitrary on a closed manifold due to global constraints. They are elementary magnetic excitations, since each element of $Z_N$ represents a unique conjugacy class. We show generalized $X$ and $Z$ operators as follows:
\begin{equation}
X = \begin{pmatrix}
0 & 1 & 0 & \cdots & 0 \\
0 & 0 & 1 & \cdots & 0 \\
0 & 0 & 0 & \cdots & 0 \\
\vdots & \vdots & \vdots & \ddots & \vdots \\
1 & 0 & 0 & \cdots & 0
\end{pmatrix}\text{, }
\qquad
Z = \begin{pmatrix}
1 & 0 & 0 & \cdots & 0 \\
0 & \omega & 0 & \cdots & 0 \\
0 & 0 & \omega^2 & \cdots & 0 \\
\vdots & \vdots & \vdots & \ddots & \vdots \\
0 & 0 & 0 & \cdots & \omega^{n-1}
\end{pmatrix}\text{,}
\end{equation}
where $\omega = e^{2\pi i / N}$ is the $N$-th root of unity. Generalized $X$ substitutes the value of a visible node $v_i$ by $v_i + 1$, utilizing the RBM’s action which operates modulo $N$.  Generalized $Z$ is implemented by adding a magnetic coupling $a = 2 \pi \mathrm{i}/N$ to the visible node. By applying these string operators composed of the generalized $X$ and $Z$, all excitations can be created on our RBM.

This method can be generalized to other lattice models with frustration-free Hamiltonians composed of two types of terms: one type is the tensor product of only Pauli-Z matrices, which constrain local flux, and the other is the tensor product of only Pauli-X matrices, which enable gauge transformations, as in all CSS codes. Applying the flux-free RBM achieves an equal superposition of all flux-free configurations, automatically satisfying the gauge transformation terms and resulting in a superposition of all logical states. For instance, this approach is applicable to the X-cube model, the checkerboard model, Haah codes, the Kitaev model associated with Abelian groups, and so on. However, the generalization to the Kitaev quantum double model associated with a non-Abelian group remains unclear.

\section{Conclusion and further work}

We analytically resolved the FRRBM proposed for the toric code, determining all possible ground states to assess the model's capabilities. We then modified this model to support arbitrary ground states through the integration of non-local connections. This enhanced model remains analytically solvable and can also be efficiently solved using machine learning techniques. We then extend our work to Kitaev quantum double model associated with abelian group $Z_N$. Our ongoing research aims to investigate feasible RBM implementations for more specific codes, including those for the Double Semion \cite{levin2005string}, Fibonacci Anyon \cite{levin2005string, lin2021generalized}, and Kitaev quantum double model associated with a non-abelian group.

\vspace{0.5cm}
\noindent\textbf{Acknowledgement.} 
The authors are partially supported by NSF grant CCF-2006667, Quantum Science Center sponsored by DOE's Office of Science, and ARO MURI. 

\bibliographystyle{unsrt}
\bibliography{ML}

\begin{thebibliography}{10}

\bibitem{kitaev2003fault}
A~Yu Kitaev.
\newblock Fault-tolerant quantum computation by anyons.
\newblock {\em Annals of Physics}, 303(1):2--30, 2003.

\bibitem{freedman2002modular}
Michael~H Freedman, Michael Larsen, and Zhenghan Wang.
\newblock A modular functor which is universal for quantum computation.
\newblock {\em Communications in Mathematical Physics}, 227(3):605--622, 2002.

\bibitem{osborne2012hamiltonian}
Tobias~J Osborne.
\newblock Hamiltonian complexity.
\newblock {\em Reports on progress in physics}, 75(2):022001, 2012.

\bibitem{hinton2006fast}
Geoffrey~E Hinton, Simon Osindero, and Yee-Whye Teh.
\newblock A fast learning algorithm for deep belief nets.
\newblock {\em Neural computation}, 18(7):1527--1554, 2006.

\bibitem{carleo2017solving}
Giuseppe Carleo and Matthias Troyer.
\newblock Solving the quantum many-body problem with artificial neural networks.
\newblock {\em Science}, 355(6325):602--606, 2017.

\bibitem{deng2017quantum}
Dong-Ling Deng, Xiaopeng Li, and S~Das Sarma.
\newblock Quantum entanglement in neural network states.
\newblock {\em Physical Review X}, 7(2):021021, 2017.

\bibitem{deng2017machine}
Dong-Ling Deng, Xiaopeng Li, and S~Das Sarma.
\newblock Machine learning topological states.
\newblock {\em Physical Review B}, 96(19):195145, 2017.

\bibitem{chen2018equivalence}
Jing Chen, Song Cheng, Haidong Xie, Lei Wang, and Tao Xiang.
\newblock Equivalence of restricted boltzmann machines and tensor network states.
\newblock {\em Physical Review B}, 97(8):085104, 2018.

\bibitem{gao2017efficient}
Xun Gao and Lu-Ming Duan.
\newblock Efficient representation of quantum many-body states with deep neural networks.
\newblock {\em Nature communications}, 8(1):662, 2017.

\bibitem{lu2019efficient}
Sirui Lu, Xun Gao, and L-M Duan.
\newblock Efficient representation of topologically ordered states with restricted boltzmann machines.
\newblock {\em Physical Review B}, 99(15):155136, 2019.

\bibitem{jia2019efficient}
Zhih-Ahn Jia, Yuan-Hang Zhang, Yu-Chun Wu, Liang Kong, Guang-Can Guo, and Guo-Ping Guo.
\newblock Efficient machine-learning representations of a surface code with boundaries, defects, domain walls, and twists.
\newblock {\em Physical Review A}, 99(1):012307, 2019.

\bibitem{van2004graphical}
Maarten Van~den Nest, Jeroen Dehaene, and Bart De~Moor.
\newblock Graphical description of the action of local clifford transformations on graph states.
\newblock {\em Physical Review A}, 69(2):022316, 2004.

\bibitem{le2008representational}
Nicolas Le~Roux and Yoshua Bengio.
\newblock Representational power of restricted boltzmann machines and deep belief networks.
\newblock {\em Neural computation}, 20(6):1631--1649, 2008.

\bibitem{huang2021neural}
Yichen Huang, Joel~E Moore, et~al.
\newblock Neural network representation of tensor network and chiral states.
\newblock {\em Physical Review Letters}, 127(17):170601, 2021.

\bibitem{liao2021graph}
Pengcheng Liao and David~L Feder.
\newblock Graph-state representation of the toric code.
\newblock {\em Physical Review A}, 104(1):012432, 2021.

\bibitem{yan2022ribbon}
Bowen Yan, Penghua Chen, and Shawn~X Cui.
\newblock Ribbon operators in the generalized kitaev quantum double model based on hopf algebras.
\newblock {\em Journal of Physics A: Mathematical and Theoretical}, 55(18):185201, 2022.

\bibitem{levin2005string}
Michael~A Levin and Xiao-Gang Wen.
\newblock String-net condensation: A physical mechanism for topological phases.
\newblock {\em Physical Review B}, 71(4):045110, 2005.

\bibitem{lin2021generalized}
Chien-Hung Lin, Michael Levin, and Fiona~J Burnell.
\newblock Generalized string-net models: A thorough exposition.
\newblock {\em Physical Review B}, 103(19):195155, 2021.

\end{thebibliography}

\newpage
\appendix
\section{Analytical solution of \texorpdfstring{$b_{f},\, w_{f,j}$}{face} in the FRRBM} \label{sec:appendix for face terms}

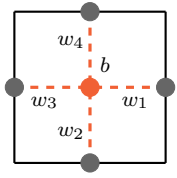
\begin{figure}[ht]
\begin{center}
\begin{tikzpicture}[scale=1.0]
\draw[very thick, dashed, RedOrange] (-2,-1) -- (0,-1);
\draw[very thick, dashed, RedOrange] (-1,0) -- (-1,-2);
\foreach \i in {-2, 0}
        \draw[thick, black] ({\i},-2) -- ({\i},0);
\foreach \j in {-2, 0}
        \draw[thick, black] (-2,{\j}) -- (0,{\j});
\foreach \j in {-2, 0}
        \filldraw[darkgray!80] (-1,{\j}) circle (0.12);
\foreach \i in {-2, 0}
        \filldraw[darkgray!80] ({\i},-1) circle (0.12);
\filldraw[RedOrange] (-1,-1) circle (0.12);
\node[black] at (-1+0.2,-1+0.3) {\scriptsize $b$};
\node[black] at (-1+0.6,-1-0.2) {\scriptsize $w_1$};
\node[black] at (-1-0.25,-1-0.6) {\scriptsize $w_2$};
\node[black] at (-1-0.6,-1-0.2) {\scriptsize $w_3$};
\node[black] at (-1-0.25,-1+0.6) {\scriptsize $w_4$};
\end{tikzpicture}
\end{center}
\caption{This lattice diagram represents a translation-invariant structure for a face-type hidden neuron, using simplified notation without the subscript $f$.}
\label{fig:face_0}
\end{figure}

To optimize $\lvert \Psi \rangle = \sum_{S} \Psi_{M}(S;\mathcal{W}) \lvert S \rangle$ to best represent the ground state $\lvert GS \rangle$, consider the following expression:
\begin{align}
    \Psi_{M}(S;\mathcal{W}) &=e^{\sum_{j}a_{j}\sigma_{j}^{z}} \prod_{v\in V} \Gamma_{v}(S;\mathcal{W}) \prod_{f\in F} \Gamma_{f}(S;\mathcal{W}),\\
    \Gamma_{v}(S;\mathcal{W})&=2\cosh(b_{v} + \sum_{j\in s(v)} w_{v,j} \sigma_{j}^{z}),\\
    \Gamma_{f}(S;\mathcal{W})&=2\cosh(b_{f} + \sum_{j\in s(f)} w_{f,j} \sigma_{j}^{z}).
\end{align}
Setting $a_{j}=0$, we treat $\lvert \Psi \rangle$ as $\lvert GS \rangle$:
\begin{align}
    \lvert GS \rangle =\sum_{S} e^{\sum_{j}a_{j}\sigma_{j}^{z}} &\prod_{v\in V} 2\cosh(b_{v} + \sum_{j\in s(v)} w_{v,j} \sigma_{j}^{z}) \nonumber \\
    &\prod_{f\in F} 2\cosh(b_{f} + \sum_{j\in s(f)} w_{f,j} \sigma_{j}^{z}) \lvert S \rangle.
\end{align}
The stabilizer condition of face operator is examined next:
\begin{equation}
B_{f} \lvert GS \rangle = \prod_{e \in s(f)} \hat{\sigma}_{e}^{z} \lvert GS \rangle = \lvert GS \rangle, \, \forall f.
\end{equation}
As the configuration $\lvert S \rangle$ remains unchanged by $\hat{\sigma}_{e}^{z}$, we get:
\begin{align}
    \prod_{e \in s(f)} \hat{\sigma}_{e}^{z} &e^{\sum_{j}a_{j}\sigma_{j}^{z}} \prod_{v\in V} \Gamma_{v}(S;\mathcal{W}) \prod_{f'\in F} \Gamma_{f'}(S;\mathcal{W}) \nonumber \\
    =&e^{\sum_{j}a_{j}\sigma_{j}^{z}} \prod_{v\in V} \Gamma_{v}(S;\mathcal{W}) \prod_{f'\in F} \Gamma_{f'}(S;\mathcal{W}), \, \forall f, \, \forall S.
\end{align}
All irrelevant terms on both sides are then cancelled:
\begin{equation}
\prod_{e \in s(f)} \hat{\sigma}_{e}^{z} \cosh(b_{f} + \sum_{j\in s(f)} w_{f,j} \sigma_{j}^{z}) = \cosh(b_{f} + \sum_{j\in s(f)} w_{f,j} \sigma_{j}^{z}), \, \forall f, \, \forall S.
\end{equation}
Due to translation invariance, it is unnecessary to repeat the calculation for all faces. Instead, the possible configurations in a single face contribute $2^{4}$ equations, as illustrated in Figure \ref{fig:face_0}:
\begin{align}
    \cosh(b-w_1+w_2&+w_3+w_4)= 0\\
    \cosh(b+w_1-w_2&-w_3-w_4)= 0\\
    \cosh(b-w_1-w_2&+w_3+w_4)\neq 0\\
    \cosh(b-w_1-w_2&-w_3-w_4)\neq 0\\
    & \vdots \nonumber
\end{align}
Solving these equations yields the complete set of solutions for the face terms $b_{f},\, w_{f,j}$: $b_{f}=0 \, (mod \, \pi)$ and $w_{f,j}=\frac{\pi}{4}i, \frac{3\pi}{4}i \, (mod \, \pi)$, where an even number of the four $w_{f,j}$ must be the same. Since the function of the face terms selectively excludes some configurations, any solution set is valid and can be chosen without loss of generality. In the main article, we choose the isotropic solution $(b_{f}, w_{f,j})=(0, \frac{\pi}{4}i)$.

\section{Analytical solution of \texorpdfstring{$b_{v},\, w_{v,j}$}{vertex} in the FRRBM} \label{sec:appendix for vertex terms}

The face terms $B_f$ typically rule out certain configurations without trivial flux, while the vertex terms $A_v$ ensures all configurations in the same logical state are uniformly weighted,  as illustrated in Figure 10. In this appendix, we continue from Equation (\ref{eqn:vertex stabilizer result}) discussed in the main article, focusing on the configurations with trivial flux illustrated in Figures \ref{tab:Selected Configuration 1} through \ref{tab:Selected Configuration 5}. We extract relevant independent equations (\ref{equ:criteria 1}) through (\ref{equ:square condition 3}) to analytically solve for $b_{v}$ and $w_{v,j}$. To simplify notation further in the calculation, we replace $\cosh$ with $\cos$ and divide all weights by $i$. We treat $b_{v}$ as a redundant parameter, similar to $a_j$, and set $b_{v}=0$, as allowing $b_{v} \in \mathbb{C}$ would introduce superfluous freedom. Further elaboration on this issue is provided at the end.

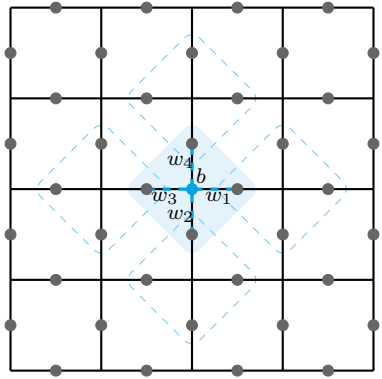
\begin{figure}[ht]
\begin{center}
\begin{tikzpicture}[scale=0.6]
\filldraw[fill=Cerulean!10, draw=none, rounded corners] (0,-1-0.5)--(1+0.5,0)--(0,1+0.5)--(-1-0.5,0)--cycle;
\draw[Cerulean!50, dashed, rounded corners] (2,-1-0.5)--(3+0.5,0)--(2,1+0.5)--(1-0.5,0)--cycle;
\draw[Cerulean!50, dashed, rounded corners] (-2,-1-0.5)--(-1+0.5,0)--(-2,1+0.5)--(-3-0.5,0)--cycle;
\draw[Cerulean!50, dashed, rounded corners] (0,1-0.5)--(1+0.5,2)--(0,3+0.5)--(-1-0.5,2)--cycle;
\draw[Cerulean!50, dashed, rounded corners] (0,-3-0.5)--(1+0.5,-2)--(0,-1+0.5)--(-1-0.5,-2)--cycle;
\foreach \i in {-4, -2, 0, 2, 4}
        \draw[thick, black] ({\i},-4) -- ({\i},4);
\foreach \j in {-4, -2, 0, 2, 4}
        \draw[thick, black] (-4,{\j}) -- (4,{\j});
\draw[very thick, dashed, Cerulean] (-1,0) -- (1,0);
\draw[very thick, dashed, Cerulean] (0,-1) -- (0,1);
\foreach \i in {-3, -1, 1, 3}
\foreach \j in {-4, -2, 0, 2, 4}
        \filldraw[darkgray!80] ({\i},{\j}) circle (0.12);
\foreach \i in {-4, -2, 0, 2, 4}
\foreach \j in {-3, -1, 1, 3}
        \filldraw[darkgray!80] ({\i},{\j}) circle (0.12);
\filldraw[Cerulean] (0,0) circle (0.12);
\node[black] at (0.2,0.3) {\scriptsize $b$};
\node[black] at (0.6,-0.2) {\scriptsize $w_1$};
\node[black] at (-0.25,-0.6) {\scriptsize $w_2$};
\node[black] at (-0.6,-0.2) {\scriptsize $w_3$};
\node[black] at (-0.25,0.6) {\scriptsize $w_4$};
\end{tikzpicture}
\end{center}
\caption{This lattice diagram represents a translation-invariant structure for a vertex-type hidden neuron, using simplified notation without the subscript $v$ and $(b_v,w_{v,j})=i*(b,w_j)$. If we flip the four qubits surrounding the central vertex, qubits contributing to the phase difference are circled for clarity.}
\label{fig:vertex_0}
\end{figure}

Equation (\ref{equ:define A}) defines the often-used phase factor $A$:
\begin{equation}\label{equ:define A}
    \cos(w_1+w_2+w_3+w_4):=A.
\end{equation}
Equation (\ref{equ:criteria 1}), the most discussed criterion, is abstracted from Figure \ref{tab:Selected Configuration 1}:
\begin{align}\label{equ:criteria 1}
    \cos(-w_1&+w_2+w_3+w_4)\cos(w_1-w_2+w_3+w_4) \nonumber \\
    &\cos(w_1+w_2-w_3+w_4)\cos(w_1+w_2+w_3-w_4)=A^4.
\end{align}
Equations (\ref{equ:square condition 1}, \ref{equ:square condition 2}, \ref{equ:square condition 3}) describe squared conditions, while Equations (\ref{equ:criteria 2}, \ref{equ:criteria 3}) specify additional criteria. All these equations are derived from the configurations shown in Figure \ref{tab:Selected Configuration 2} through \ref{tab:Selected Configuration 5}:
\begin{align}\label{equ:square condition 1}
    [\cos(-w_1&-w_2+w_3+w_4)\cos(w_1+w_2-w_3-w_4)]^L=A^{2L}\text{\, for any $L$} \nonumber \\
    &\Rightarrow \cos^2(-w_1-w_2+w_3+w_4)=A^2.
\end{align}
\begin{align}\label{equ:square condition 2}
    [\cos(-w_1&+w_2+w_3-w_4)\cos(w_1-w_2-w_3+w_4)]^L=A^{2L}\text{\, for any $L$} \nonumber \\
    &\Rightarrow \cos^2(w_1-w_2-w_3+w_4)=A^2.
\end{align}
\begin{align}
    \cos(\!-w_1\!&+w_2\!-w_3\!+w_4)\cos(\!-w_1\!+w_2\!+w_3\!+w_4)\cos(w_1\!+w_2\!-w_3\!+w_4)=A^3 \nonumber \\
    \cos(w_1\!&-w_2\!+w_3\!-w_4)\cos(w_1\!-w_2\!+w_3\!+w_4)\cos(w_1\!+w_2\!+w_3\!-w_4)=A^3 \nonumber\\
    &\Rightarrow \cos^2(w_1-w_2+w_3-w_4)=A^2. \label{equ:square condition 3}\\
    &\Rightarrow \cos^2(w_1-w_2+w_3+w_4)\cos^2(w_1+w_2+w_3-w_4)=A^4. \label{equ:criteria 2} \\
    &\Rightarrow \cos^2(-w_1+w_2+w_3+w_4)\cos^2(w_1+w_2-w_3+w_4)=A^4. \label{equ:criteria 3}
\end{align}
Next, we need to solve and discuss Equations (\ref{equ:define A}) through (\ref{equ:criteria 3}):
\begin{align}
    &\text{Equation (\ref{equ:define A})+(\ref{equ:square condition 1}):\quad} w_1+w_2=0 \text{\, or \,} w_3+w_4=0 \nonumber \\
    &\text{Equation (\ref{equ:define A})+(\ref{equ:square condition 2}):\quad} w_1+w_4=0 \text{\, or \,} w_2+w_3=0 \nonumber \\
    &\text{Equation (\ref{equ:define A})+(\ref{equ:square condition 3}):\quad} w_1+w_3=0 \text{\, or \,} w_2+w_4=0 \nonumber \\
    &\Rightarrow -w_1=w_2=w_3=w_4 \, (\text{mod}\, \frac{\pi}{2}) \,\text{and alternations.}  \label{equ:scenario A} \\
    &\text{or}\quad w_1=w_2=w_3=0 \, \text{or} \, \frac{\pi}{4} \, (\text{mod}\, \frac{\pi}{2}) \,\text{and alternations.} \label{equ:scenario B}
\end{align}

In the first scenario derived in Equation (\ref{equ:scenario A}), without loss of generality, we can set $w_1=-w+\frac{\pi}{2}m_1$, $w_2=w+\frac{\pi}{2}m_2$, $w_3=w+\frac{\pi}{2}m_3$, and $w_4=w+\frac{\pi}{2}m_4$, where $m_1$, $m_2$, $m_3$, $m_4$ $\in \mathbb{N}$. Subsequently, the criteria in Equation (\ref{equ:criteria 1}) is rewritten as
\begin{align}
    \cos[4w&+\frac{\pi}{2}(M-2m_1)]\cos[\frac{\pi}{2}(M-2m_2)]\nonumber \\
    &\cos[\frac{\pi}{2}(M-2m_3)]\cos[\frac{\pi}{2}(M-2m_4)]=\cos^4(2w+\frac{\pi}{2}M), \label{equ:result 1}
\end{align}
where $M:=m_1+m_2+m_3+m_4$. And $\cos[\frac{\pi}{2}(M-2m_2)]\neq0$ $\,\Rightarrow\,$ $M$ is even, thus the terms with $\frac{\pi}{2}$ above could be rearranged as follows:
\begin{align*}
    \cos[4w\!+\!\frac{\pi}{2}(M\!-\!2m_1)]&=\cos(4w)\cos\left[\frac{\pi}{2}(M\!-\!2m_1)\right]\!-\!\sin(4w)\sin\left[\frac{\pi}{2}(M\!-\!2m_1)\right]\\
    &=\cos(4w)\cos\left[\frac{\pi}{2}(M\!-\!2m_1)\right],
\end{align*}
or vice versa:
\begin{equation*}
    \cos[\frac{\pi}{2}(M-2m_2)]\cos[\frac{\pi}{2}(M-2m_3)]=\cos[\frac{\pi}{2}(2M-2m_2-2m_3)].
\end{equation*}
Then Equation (\ref{equ:result 1}) is simplified as
\begin{align}
    \cos(4w)\cos[\frac{\pi}{2}(4M-2M)]&=[\cos(2w)\cos(\frac{\pi}{2}M)]^4 \nonumber \\
    \cos(4w)&=\cos^4(2w). \label{equ:result 1+}
\end{align}
Solving Equation (\ref{equ:result 1+}), we get $\cos(4w)=1$, implying $w=0\,(\text{mod}\,\frac{\pi}{2})$. Considering the definition of $A$ in Equation (\ref{equ:define A}), which is not $0$, we conclude: \textbf{even number of w are $\boldsymbol{0\,(\text{mod}\,\pi)}$ and the others are $\boldsymbol{\frac{\pi}{2}\,(\text{mod}\,\pi)}$. For these solutions, $\boldsymbol{A}$ is either $\boldsymbol{+1}$ or $\boldsymbol{-1}$.}

In the second scenario derived in Equation (\ref{equ:scenario B}), without loss of generality, we can set $w_1=w_2=w_3=w$, $w=0$ or $\frac{\pi}{4}\,(\text{mod}\,\frac{\pi}{2})$ and $w_4$ is free. By inserting all possible values of $w_1$, $w_2$ and $w_3$ into Equations (\ref{equ:criteria 1}) and (\ref{equ:criteria 2}, \ref{equ:criteria 3}), we can find the allowed solutions: \textbf{three of w equal to $\boldsymbol{0 \, (\text{mod} \, \frac{\pi}{2})}$ and the other serves as a normalization parameter such that $ \boldsymbol{\cos(3w + w_4) = A}$.}

In the calculation above, we treat $b$ as a redundant parameter and set $b=0$. Reader can practice by setting $b=\frac{\pi}{2}$ to obtain another set of solutions, which yields results similar to those above. To understand how allowing $b \in \mathbb{C}$ introduces superfluous freedom, one can abstract Equation (\ref{equ:extra condition}) from Figure \ref{tab:Extra Configuration} and incorporate $b$ into Equation (\ref{equ:define A}). By re-deriving Equations (\ref{equ:criteria 1}) through (\ref{equ:square condition 3}) and substituting them into Equation (\ref{equ:extra condition}), one can obtain Equation (\ref{equ:restriction on b}), which represents the restriction on $b$. There are infinite many possibilities, and above two learnable solutions emerge through the training process introduced in the next section.
\begin{align}
    \cos(b\!-\!w_1\!&-\!w_2\!+\!w_3\!+\!w_4)\cos(b\!+\!w_1\!-\!w_2\!-\!w_3\!+\!w_4)\cos(b\!+\!w_1\!+\!w_2\!-\!w_3\!-\!w_4) \nonumber \\
    \cos(b\!-\!w_1\!&+\!w_2\!+\!w_3\!-\!w_4)\cos^{2}(b\!+\!w_1\!-\!w_2\!-\!w_3\!-\!w_4)\cos^{2}(b\!-\!w_1\!+\!w_2\!-\!w_3\!-\!w_4) \nonumber \\
    &\cos^{2}(b\!-\!w_1\!-\!w_2\!+\!w_3\!-\!w_4)\cos^{2}(b\!-\!w_1\!-\!w_2\!-\!w_3\!+\!w_4)=A^{12}. \label{equ:extra condition}
\end{align}
\begin{equation}\label{equ:restriction on b}
    \cos^{4}(b-w_1-w_2-w_3-w_4) = \cos^{4}(b+w_1+w_2+w_3+w_4).
\end{equation}

\begin{table}[ht]
\centering
\begin{tabular}{m{5.5cm} | m{5.5cm}} 
\begin{center}
\begin{tikzpicture}[scale=0.6]
\foreach \i in {-4, -2, 0, 2, 4}
        \draw[thick, black] ({\i},-4) -- ({\i},4);
\foreach \j in {-4, -2, 0, 2, 4}
        \draw[thick, black] (-4,{\j}) -- (4,{\j});
\foreach \i in {-3, -1, 1, 3}
\foreach \j in {-4, -2, 0, 2}
        \filldraw[darkgray!80] ({\i},{\j}) circle (0.12);
\foreach \i in {-2, 0, 2, 4}
\foreach \j in {-3, -1, 1, 3}
        \filldraw[darkgray!80] ({\i},{\j}) circle (0.12);

\node[black] at (0+0.3,0-0.3) {\footnotesize $v_{0}$};
\end{tikzpicture}
\end{center}
&
\begin{center}
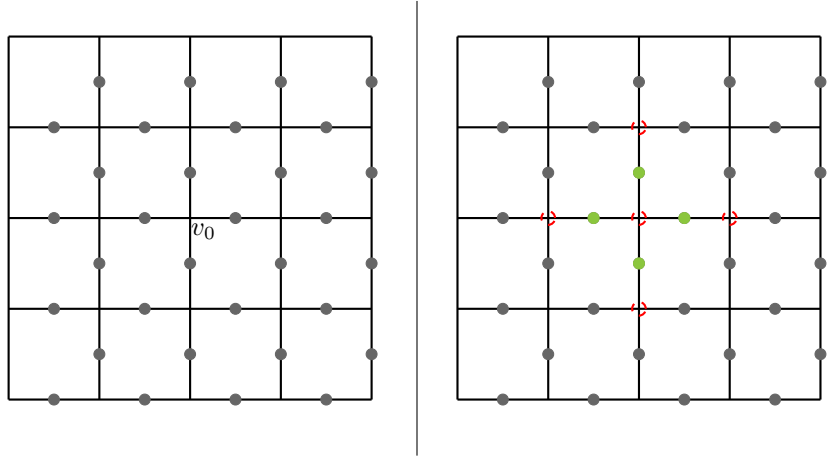

\begin{tikzpicture}[scale=0.6]
\foreach \i in {-4, -2, 0, 2, 4}
        \draw[thick, black] ({\i},-4) -- ({\i},4);
\foreach \j in {-4, -2, 0, 2, 4}
        \draw[thick, black] (-4,{\j}) -- (4,{\j});
\foreach \i in {-3, -1, 1, 3}
\foreach \j in {-4, -2, 0, 2}
        \filldraw[darkgray!80] ({\i},{\j}) circle (0.12);
\foreach \i in {-2, 0, 2, 4}
\foreach \j in {-3, -1, 1, 3}
        \filldraw[darkgray!80] ({\i},{\j}) circle (0.12);

\foreach \i in {-1, 1}
\filldraw[LimeGreen] ({\i},0) circle (0.12);
\foreach \j in {-1, 1}
\filldraw[LimeGreen] (0,{\j}) circle (0.12);
\foreach \i in {-2, 0, 2}
\draw[dashed, thick, red] ({\i},0) circle (0.15);
\foreach \j in {-2, 2}
\draw[dashed, thick, red] (0,{\j}) circle (0.15);
\end{tikzpicture}
\end{center}
\end{tabular}
\captionof{figure}[foo]{In this configuration, each black dot represents a qubit in state $\lvert -1 \rangle$, while each green dot indicates a qubit in state $\lvert +1 \rangle$. Starting with the initial configuration on the left, a vertex operator is applied at vertex $v_0$ to flip the adjacent four qubits. The resulting configuration, displayed on the right, features five vertices encircled in red that contribute to the phase difference. By comparing the phase contributions from these vertices in both configurations, we derive Equation (\ref{equ:criteria 1}), which is a crucial criterion for our calculation.} 
\label{tab:Selected Configuration 1}
\end{table}

\begin{table}[ht]
\centering
\begin{tabular}{m{5.5cm} | m{5.5cm}} 
\begin{center}
\begin{tikzpicture}[scale=0.6]
\foreach \i in {-4, -2, 0, 2, 4}
        \draw[thick, black] ({\i},-4) -- ({\i},4);
\foreach \j in {-4, -2, 0, 2, 4}
        \draw[thick, black] (-4,{\j}) -- (4,{\j});
\foreach \i in {-3, -1, 1, 3}
\foreach \j in {-4, -2, 0, 2}
        \filldraw[darkgray!80] ({\i},{\j}) circle (0.12);
\foreach \i in {-2, 0, 2, 4}
\foreach \j in {-3, -1, 1, 3}
        \filldraw[darkgray!80] ({\i},{\j}) circle (0.12);

\node[black] at (-0.45-2,-0.3-2) {\footnotesize $v_{i-1}$};
\node[black] at (-0.3,-0.3) {\footnotesize $v_{i}$};
\node[black] at (2-0.45,2-0.3) {\footnotesize $v_{i+1}$};

\draw[dotted, thick, red] (-4,-3) -- (3,4);
\draw[dotted, thick, red] (-3,-4) -- (4,3);
\end{tikzpicture}
\end{center}
&
\begin{center}
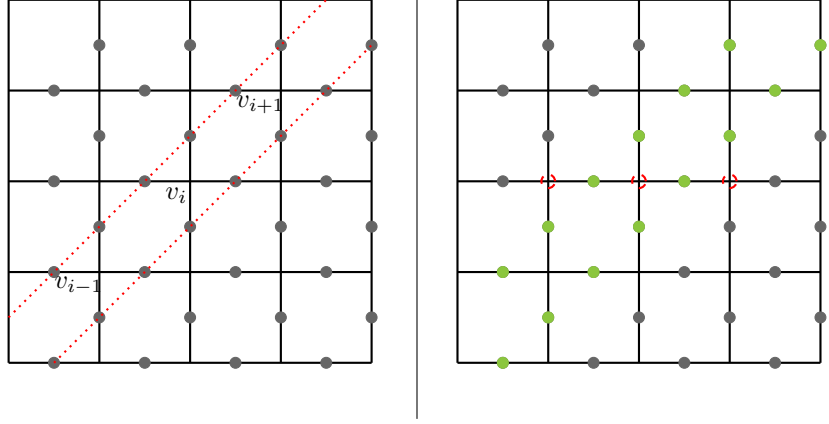

\begin{tikzpicture}[scale=0.6]
\foreach \i in {-4, -2, 0, 2, 4}
        \draw[thick, black] ({\i},-4) -- ({\i},4);
\foreach \j in {-4, -2, 0, 2, 4}
        \draw[thick, black] (-4,{\j}) -- (4,{\j});
\foreach \i in {-3, -1, 1, 3}
\foreach \j in {-4, -2, 0, 2}
        \filldraw[darkgray!80] ({\i},{\j}) circle (0.12);
\foreach \i in {-2, 0, 2, 4}
\foreach \j in {-3, -1, 1, 3}
        \filldraw[darkgray!80] ({\i},{\j}) circle (0.12);

\foreach \i in {-3,-2,-1,0,1,2}
        \filldraw[LimeGreen] ({\i},{\i+1}) circle (0.12);
\foreach \i in {-3,-2,-1,0,1,2,3,4}
        \filldraw[LimeGreen] ({\i},{\i-1}) circle (0.12);
\foreach \i in {-2, 0, 2}
        \draw[dashed, thick, red] ({\i},0) circle (0.15);
\end{tikzpicture}
\end{center}
\end{tabular}
\captionof{figure}[foo]{Starting with the initial configuration illustrated on the left, we apply vertex operators diagonally at vertices $v_0$. The resultant configuration, showcased on the right, exhibits translational symmetry horizontally. Notably, the three vertices encircled in red contribute to the phase difference. By analyzing the phase contributions from these three vertices in both configurations, we deduce Equation (\ref{equ:square condition 1}). This equation represents one of the three pivotal square conditions essential for our calculation.} 
\label{tab:Selected Configuration 2}
\end{table}

\begin{table}[ht]
\centering
\begin{tabular}{m{5.5cm} | m{5.5cm}} 
\begin{center}
\begin{tikzpicture}[scale=0.6]
\foreach \i in {-4, -2, 0, 2, 4}
        \draw[thick, black] ({\i},-4) -- ({\i},4);
\foreach \j in {-4, -2, 0, 2, 4}
        \draw[thick, black] (-4,{\j}) -- (4,{\j});
\foreach \i in {-3, -1, 1, 3}
\foreach \j in {-4, -2, 0, 2}
        \filldraw[darkgray!80] ({\i},{\j}) circle (0.12);
\foreach \i in {-2, 0, 2, 4}
\foreach \j in {-3, -1, 1, 3}
        \filldraw[darkgray!80] ({\i},{\j}) circle (0.12);

\node[black] at (0.5-2,-0.3+2) {\footnotesize $v_{i-1}$};
\node[black] at (0.3,-0.3) {\footnotesize $v_{i}$};
\node[black] at (2+0.5,-2-0.3) {\footnotesize $v_{i+1}$};
\draw[dotted, thick, red] (4,-3) -- (-3,4);
\draw[dotted, thick, red] (3,-4) -- (-4,3);
\end{tikzpicture}
\end{center}
&
\begin{center}
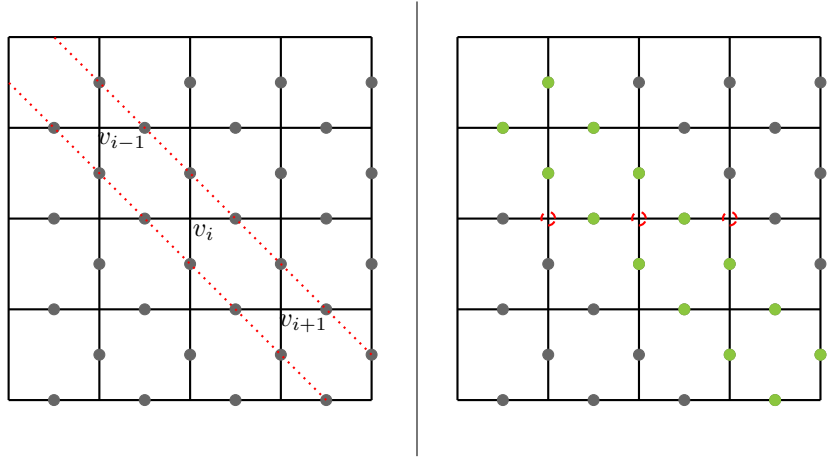

\begin{tikzpicture}[scale=0.6]
\foreach \i in {-4, -2, 0, 2, 4}
        \draw[thick, black] ({\i},-4) -- ({\i},4);
\foreach \j in {-4, -2, 0, 2, 4}
        \draw[thick, black] (-4,{\j}) -- (4,{\j});
\foreach \i in {-3, -1, 1, 3}
\foreach \j in {-4, -2, 0, 2}
        \filldraw[darkgray!80] ({\i},{\j}) circle (0.12);
\foreach \i in {-2, 0, 2, 4}
\foreach \j in {-3, -1, 1, 3}
        \filldraw[darkgray!80] ({\i},{\j}) circle (0.12);

\foreach \i in {-2,-1,0,1,2,3,4}
        \filldraw[LimeGreen] ({\i},{-\i+1}) circle (0.12);
\foreach \i in {-3,-2,-1,0,1,2,3}
        \filldraw[LimeGreen] ({\i},{-\i-1}) circle (0.12);
\foreach \i in {-2, 0, 2}
        \draw[dashed, thick, red] ({\i},0) circle (0.15);
\end{tikzpicture}
\end{center}
\end{tabular}
\captionof{figure}[foo]{Similarly, applying vertex operators diagonally in the perpendicular direction, we obtain Equation (\ref{equ:square condition 2}), another pivotal square condition.} 
\label{tab:Selected Configuration 3}
\end{table}

\begin{table}[ht]
\centering
\begin{tabular}{m{5.5cm} | m{5.5cm}} 
\begin{center}
\begin{tikzpicture}[scale=0.6]
\foreach \i in {-4, -2, 0, 2, 4}
        \draw[thick, black] ({\i},-4) -- ({\i},4);
\foreach \j in {-4, -2, 0, 2, 4}
        \draw[thick, black] (-4,{\j}) -- (4,{\j});
\foreach \i in {-3, -1, 1, 3}
\foreach \j in {-4, -2, 0, 2}
        \filldraw[darkgray!80] ({\i},{\j}) circle (0.12);
\foreach \i in {-2, 0, 2, 4}
\foreach \j in {-3, -1, 1, 3}
        \filldraw[darkgray!80] ({\i},{\j}) circle (0.12);

\node[black] at (0.45,-2-0.3) {\footnotesize $v_{i-1}$};
\node[black] at (0.3,-0.3) {\footnotesize $v_{i}$};
\node[black] at (0.45,2-0.3) {\footnotesize $v_{i+1}$};
\draw[dotted, thick, red] (-1,-4) -- (-1,4);
\draw[dotted, thick, red] (1,-4) -- (1,4);
\end{tikzpicture}
\end{center}
&
\begin{center}
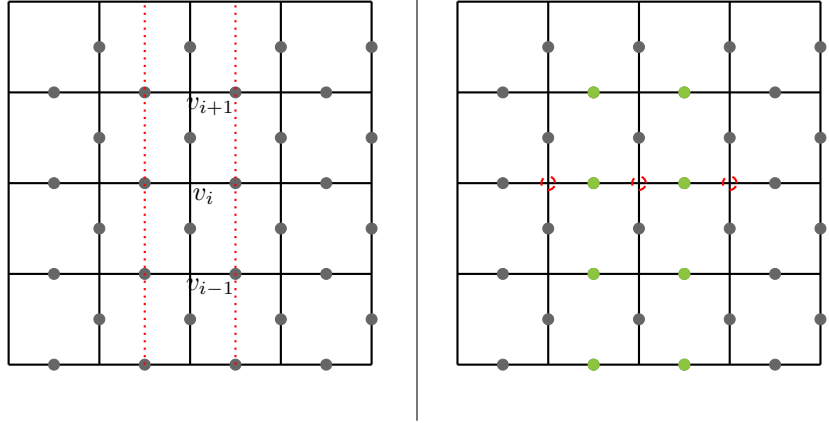

\begin{tikzpicture}[scale=0.6]
\foreach \i in {-4, -2, 0, 2, 4}
        \draw[thick, black] ({\i},-4) -- ({\i},4);
\foreach \j in {-4, -2, 0, 2, 4}
        \draw[thick, black] (-4,{\j}) -- (4,{\j});
\foreach \i in {-3, -1, 1, 3}
\foreach \j in {-4, -2, 0, 2}
        \filldraw[darkgray!80] ({\i},{\j}) circle (0.12);
\foreach \i in {-2, 0, 2, 4}
\foreach \j in {-3, -1, 1, 3}
        \filldraw[darkgray!80] ({\i},{\j}) circle (0.12);

\foreach \i in {-1,1}
\foreach \j in {-4,-2,0,2}
        \filldraw[LimeGreen] ({\i},{\j}) circle (0.12);
\foreach \i in {-2, 0, 2}
        \draw[dashed, thick, red] ({\i},0) circle (0.15);
\end{tikzpicture}
\end{center}
\end{tabular}
\captionof{figure}[foo]{This configuration corresponds to the first equation to derive the Equation (\ref{equ:square condition 3}), the last pivotal square condition.} 
\label{tab:Selected Configuration 4}
\end{table}

\begin{table}[ht]
\centering
\begin{tabular}{m{5.5cm} | m{5.5cm}} 
\begin{center}
\begin{tikzpicture}[scale=0.6]
\foreach \i in {-4, -2, 0, 2, 4}
        \draw[thick, black] ({\i},-4) -- ({\i},4);
\foreach \j in {-4, -2, 0, 2, 4}
        \draw[thick, black] (-4,{\j}) -- (4,{\j});
\foreach \i in {-3, -1, 1, 3}
\foreach \j in {-4, -2, 0, 2}
        \filldraw[darkgray!80] ({\i},{\j}) circle (0.12);
\foreach \i in {-2, 0, 2, 4}
\foreach \j in {-3, -1, 1, 3}
        \filldraw[darkgray!80] ({\i},{\j}) circle (0.12);

\node[black] at (-2+0.45,-0.3) {\footnotesize $v_{i-1}$};
\node[black] at (0.3,-0.3) {\footnotesize $v_{i}$};
\node[black] at (2+0.45,-0.3) {\footnotesize $v_{i+1}$};

\draw[dotted, thick, red] (-4,1) -- (4,1);
\draw[dotted, thick, red] (-4,-1) -- (4,-1);
\end{tikzpicture}
\end{center}
&
\begin{center}
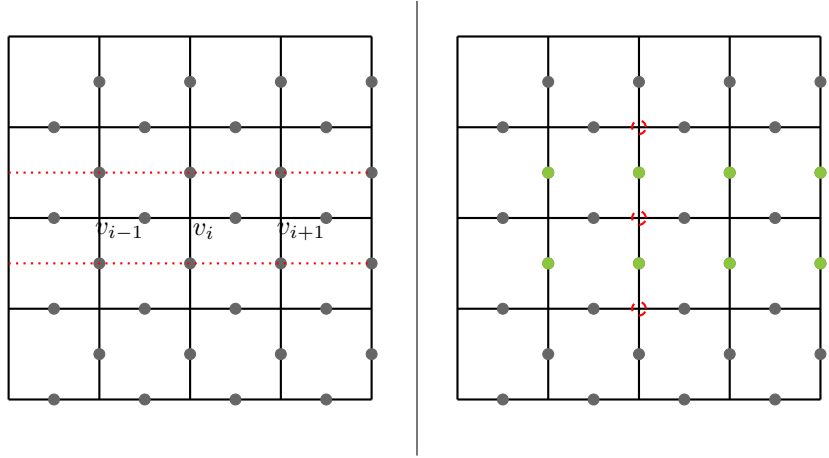

\begin{tikzpicture}[scale=0.6]
\foreach \i in {-4, -2, 0, 2, 4}
        \draw[thick, black] ({\i},-4) -- ({\i},4);
\foreach \j in {-4, -2, 0, 2, 4}
        \draw[thick, black] (-4,{\j}) -- (4,{\j});
\foreach \i in {-3, -1, 1, 3}
\foreach \j in {-4, -2, 0, 2}
        \filldraw[darkgray!80] ({\i},{\j}) circle (0.12);
\foreach \i in {-2, 0, 2, 4}
\foreach \j in {-3, -1, 1, 3}
        \filldraw[darkgray!80] ({\i},{\j}) circle (0.12);

\foreach \i in {-2,0,2,4}
\foreach \j in {-1,1}
        \filldraw[LimeGreen] ({\i},{\j}) circle (0.12);
\foreach \j in {-2, 0, 2}
        \draw[dashed, thick, red] (0,{\j}) circle (0.15);
\end{tikzpicture}
\end{center}
\end{tabular}
\captionof{figure}[foo]{This configuration corresponds to the second equation to derive the Equation (\ref{equ:square condition 3}), the last pivotal square condition.} 
\label{tab:Selected Configuration 5}
\end{table}

\begin{table}[ht]
\centering
\begin{tabular}{m{5.5cm} | m{5.5cm}} 
\begin{center}
\begin{tikzpicture}[scale=0.6]
\foreach \i in {-4, -2, 0, 2, 4}
        \draw[thick, black] ({\i},-4) -- ({\i},4);
\foreach \j in {-4, -2, 0, 2, 4}
        \draw[thick, black] (-4,{\j}) -- (4,{\j});
\foreach \i in {-3, -1, 1, 3}
\foreach \j in {-4, -2, 0, 2}
        \filldraw[darkgray!80] ({\i},{\j}) circle (0.12);
\foreach \i in {-2, 0, 2, 4}
\foreach \j in {-3, -1, 1, 3}
        \filldraw[darkgray!80] ({\i},{\j}) circle (0.12);
\node[black] at (0.3,-0.3) {\footnotesize $v_1$};
\node[black] at (2.3,-0.3) {\footnotesize $v_2$};
\node[black] at (0.3,-2.3) {\footnotesize $v_3$};
\node[black] at (2.3,-2.3) {\footnotesize $v_4$};
\end{tikzpicture}
\end{center}
&
\begin{center}
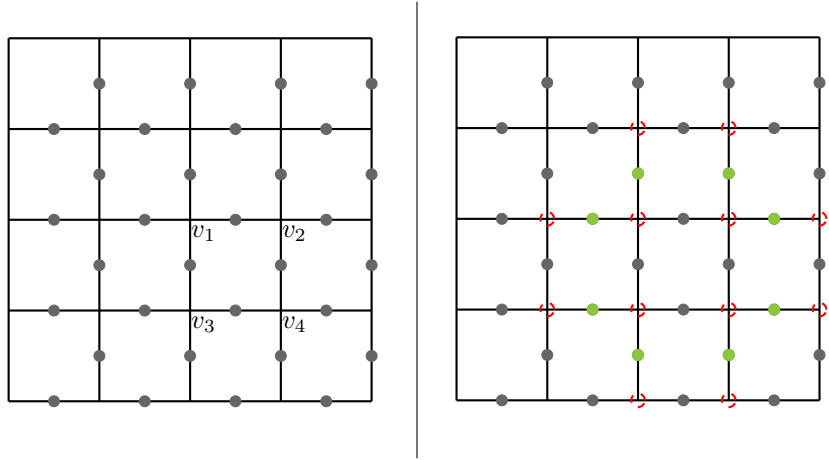

\begin{tikzpicture}[scale=0.6]
\foreach \i in {-4, -2, 0, 2, 4}
        \draw[thick, black] ({\i},-4) -- ({\i},4);
\foreach \j in {-4, -2, 0, 2, 4}
        \draw[thick, black] (-4,{\j}) -- (4,{\j});
\foreach \i in {-3, -1, 1, 3}
\foreach \j in {-4, -2, 0, 2}
        \filldraw[darkgray!80] ({\i},{\j}) circle (0.12);
\foreach \i in {-2, 0, 2, 4}
\foreach \j in {-3, -1, 1, 3}
        \filldraw[darkgray!80] ({\i},{\j}) circle (0.12);

\foreach \i in {-1, 3}
\foreach \j in {-2, 0}
        \filldraw[LimeGreen] ({\i},{\j}) circle (0.12);
\foreach \i in {0, 2}
\foreach \j in {-3, 1}
        \filldraw[LimeGreen] ({\i},{\j}) circle (0.12);
\foreach \i in {-2, 0, 2, 4}
\foreach \j in {-2, 0}
        \draw[dashed, thick, red] ({\i},{\j}) circle (0.15);
\foreach \i in {0, 2}
\foreach \j in {-4, 2}
        \draw[dashed, thick, red] ({\i},{\j}) circle (0.15);
\end{tikzpicture}
\end{center}
\end{tabular}
\captionof{figure}[foo]{This configuration corresponds to Equation (\ref{equ:extra condition}), the condition to find the restriction on $b_v$.} 
\label{tab:Extra Configuration}
\end{table}
 
\section{Machine Learning of the FRRBM} \label{sec:appendix for FRRBM}

\begin{figure}[ht]
  \centering
  \begin{minipage}[b]{0.48\textwidth}
    \centering
    \includegraphics[width=\textwidth]{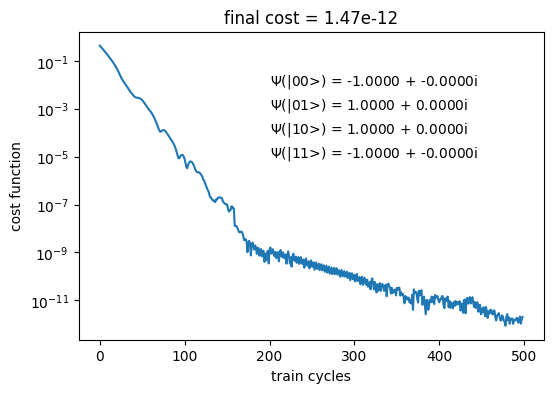}
  \end{minipage}
  \hfill
  \begin{minipage}[b]{0.48\textwidth}
    \centering
    \includegraphics[width=\textwidth]{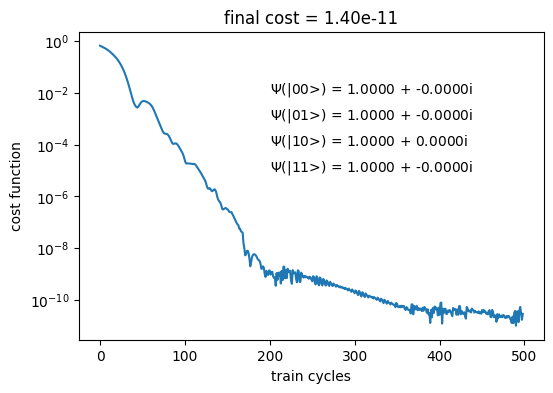}
  \end{minipage}
  \caption{On a $3\times3$ lattice, the left plot shows the training result for the isotropic setting $w_{v,j} = \frac{\pi}{2}i$, and right for the anisotropic setting $w_{v,j}=0,0,\frac{\pi}{2}i,\frac{\pi}{2}i$.}
  \label{fig:FRRBM different sizes 3x3}
\end{figure}

\begin{figure}[ht]
  \centering
  \begin{minipage}[b]{0.48\textwidth}
    \centering
    \includegraphics[width=\textwidth]{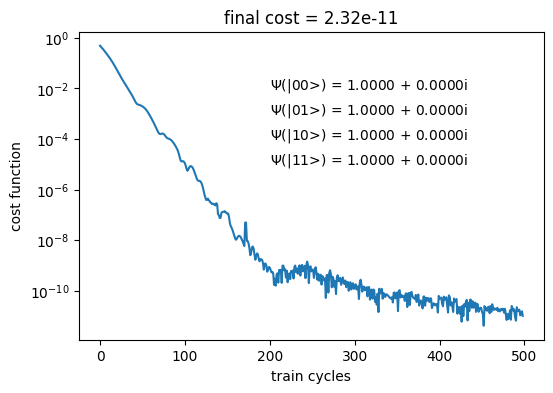}
  \end{minipage}
  \hfill
  \begin{minipage}[b]{0.48\textwidth}
    \centering
    \includegraphics[width=\textwidth]{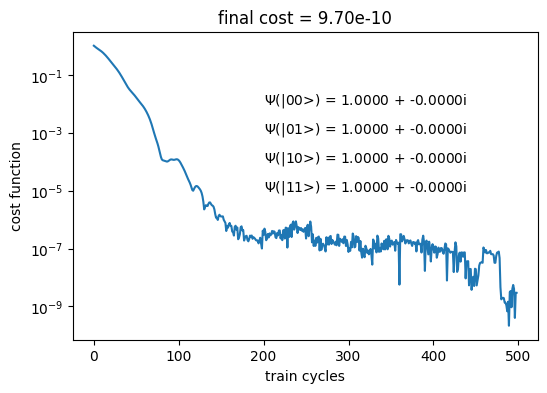}
  \end{minipage}
  \caption{On a $4\times4$ lattice, the left plot shows the training result for the isotropic setting $w_{v,j} = \frac{\pi}{2}i$, and right for the anisotropic setting $w_{v,j}=0,0,\frac{\pi}{2}i,\frac{\pi}{2}i$. Despite the varied interaction settings, both configurations yield identical ground states. This outcome contrasts with the results shown in Figure \ref{fig:FRRBM different sizes 3x3}, where the ground states differ significantly.}
  \label{fig:FRRBM different sizes 4x4}
\end{figure}

\begin{figure}[ht]
  \centering
  \begin{minipage}[b]{0.48\textwidth}
    \centering
    \includegraphics[width=\textwidth]{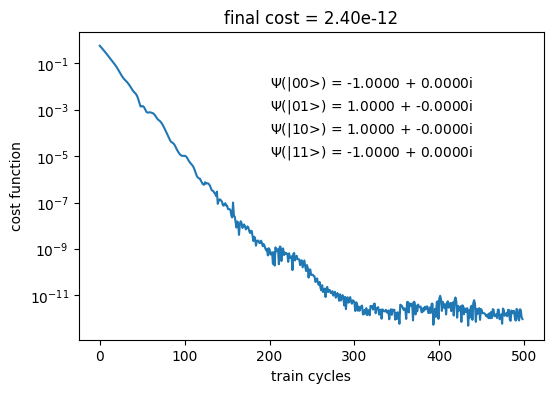}
  \end{minipage}
  \hfill
  \begin{minipage}[b]{0.48\textwidth}
    \centering
    \includegraphics[width=\textwidth]{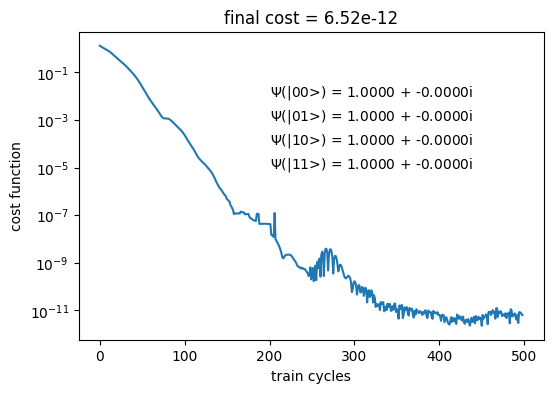}
  \end{minipage}
  \caption{On a $5\times5$ lattice, the left plot shows the training result for the isotropic setting $w_{v,j} = \frac{\pi}{2}i$, and right for the anisotropic setting $w_{v,j}=0,0,\frac{\pi}{2}i,\frac{\pi}{2}i$. This outcome align with the results shown in Figure \ref{fig:FRRBM different sizes 3x3}.}
  \label{fig:FRRBM different sizes 5x5}
\end{figure}

To further elucidate the analytical solutions derived in the main article for the FRRBM illustrated in Figure \ref{tab:torus FRRBM}, we numerically determine the ground state solution from Equations (\ref{eqn:training target}) and (\ref{eqn:FRRBM wave function}) by applying a vertex stabilizer condition on square lattices of various sizes. Namleluy, Figures \ref{fig:FRRBM different sizes 3x3}, \ref{fig:FRRBM different sizes 4x4} and \ref{fig:FRRBM different sizes 5x5} show their factorization on $3 \times 3$, $4 \times 4$ and $5 \times 5$ square lattices with different initial settings, where the common setting is $(a_j,b_f,w_{f,j},b_v)=(0,0,\frac{\pi}{4}i,0)$, with variations in $w_{v,j}$ making the difference.

As the system size increases, achieving the same level of precision necessitates a corresponding increase in the number of selected configurations. For example, in the plots shown above, the numbers of selected configurations are 50, 100, and 150 for progressively larger systems. Consequently, the training time and computational resources required increase proportionally with the system size. This escalation represents a common challenge when optimizing parameters in neural network representations of quantum states.

\section{Machine Learning of the RBM} \label{sec:appendix for abitrary}

Illustrated in Figure \ref{tab:arbitrary torus RBM}, we pick the isotropic setting $(a_j,b_f,w_{f,j},b_v,w_{v,j})=(0,0,\frac{\pi}{4}i,0,\frac{\pi}{2}i)$ and uniformly weighted every new connection ($w_{x,y,z}=\frac{\pi}{4}i$). Then three hidden neurons ($h_x$, $h_y$, $h_z$) are introduced into the FRRBM to simulate an arbitrary ground state. Reader can verify that the inclusion of $h_x$ and $h_y$ (inspired by the logical operators $Z_v$ and $Z_h$) allows for the simulation of any specific degeneracy state, while $h_z$ enables the representation of any arbitrary ground state as a linear combination within the degeneracy basis. Then, using Equations from (\ref{eqn:ratio_00}) to (\ref{eqn:ratio_11}), we can analytically solve the weights for arbitrary ground state.

On the other hand, illustrated in Figure \ref{tab:basis in the RBM}, the set of selected configurations $\mathcal{S'}=\{$ $S_1$, $S_2$, $S_3$, $S_4 \}$ are chosen from the equi-positioned configurations of the states $\lvert 00 \rangle$, $\lvert 01 \rangle$, $\lvert 10 \rangle$, $\lvert 11 \rangle$, respectively. Employing the condition $\langle GS \lvert S1 \rangle\!:\!\langle GS \lvert S2 \rangle\!:\!\langle GS \lvert S3 \rangle\!:\!\langle GS \lvert S4 \rangle = \langle GS \lvert 00 \rangle\!:\!\langle GS \lvert 01 \rangle\!:\!\langle GS \lvert 10 \rangle\!:\!\langle GS \lvert 11 \rangle $, we can numerically train the weights for arbitrary ground state according to the ratio conditions.

\begin{table}[ht]
\centering
\begin{tabular}{m{5.5cm}  m{5.5cm}} 
\centering
\begin{tikzpicture}[scale=0.65]

\filldraw[very thick, dashed, OliveGreen] (3,-2) -- (3,4);
\filldraw[very thick, dashed, brown] (-2,1) -- (4,1);
\foreach \i in {0, 2}
        \draw[thick, black] ({\i},-2) -- ({\i},4);
\foreach \j in {0, 2}
        \draw[thick, black] (-2,{\j}) -- (4,{\j});
\foreach \i in {-2, 4}
    \draw[very thick, MidnightBlue, postaction={decorate, decoration={markings,
    mark=at position 0.45 with {\arrow[MidnightBlue]{stealth}}}}]
    ({\i},-2) -- ({\i},4);
\foreach \j in {-2, 4}
    \draw[very thick, BrickRed, postaction={decorate, decoration={markings,
    mark=at position 0.45 with {\arrow[BrickRed]{stealth}}}}]
    (-2,{\j}) -- (4,{\j});
\foreach \i in {-1, 1, 3}
\foreach \j in {-2, 0, 2}
        \filldraw[darkgray!80] ({\i},{\j}) circle (0.1);
\foreach \i in {0, 2, 4}
\foreach \j in {-1, 1, 3}
        \filldraw[darkgray!80] ({\i},{\j}) circle (0.1);
\node[black] at (4.5,1) {\scriptsize $X_{h}$};
\node[black] at (3,-2.5) {\scriptsize $X_{v}$};
\end{tikzpicture}
&
\centering
\begin{tikzpicture}[scale=0.65]
\foreach \i in {0, 2}
        \draw[thick, black] ({\i},-2) -- ({\i},4);
\foreach \j in {0, 2}
        \draw[thick, black] (-2,{\j}) -- (4,{\j});
\foreach \i in {-2, 4}
    \draw[very thick, MidnightBlue, postaction={decorate, decoration={markings,
    mark=at position 0.45 with {\arrow[MidnightBlue]{stealth}}}}]
    ({\i},-2) -- ({\i},4);
\foreach \j in {-2, 4}
    \draw[very thick, BrickRed, postaction={decorate, decoration={markings,
    mark=at position 0.45 with {\arrow[BrickRed]{stealth}}}}]
    (-2,{\j}) -- (4,{\j});
\foreach \i in {-1, 1, 3}
        \filldraw[very thick, dashed, Orchid] (5,-3) -- (\i,-2);
\foreach \j in {-1, 1, 3}
        \filldraw[very thick, dashed, Orchid] (5,-3) -- (4,\j);
\foreach \i in {-1, 1, 3}
        \filldraw[very thick, dashed, OliveGreen] (1,-3) -- (\i,-2);
\foreach \j in {-1, 1, 3}
        \filldraw[very thick, dashed, brown] (5,1) -- (4,\j);
\foreach \i in {-1, 1, 3}
\foreach \j in {-2, 0, 2}
        \filldraw[darkgray!80] ({\i},{\j}) circle (0.1);
\foreach \i in {0, 2, 4}
\foreach \j in {-1, 1, 3}
        \filldraw[darkgray!80] ({\i},{\j}) circle (0.1);
\filldraw[OliveGreen!70] (1,-3) circle (0.2);
\filldraw[brown!80] (5,1) circle (0.2);
\filldraw[Orchid!80] (5,-3) circle (0.2);

\node[black] at (1+0.5,-3) {\scriptsize $b_{x}$};
\node[black] at (5+0.5,1) {\scriptsize $b_{y}$};
\node[black] at (5+0.5,-3) {\scriptsize $b_{z}$};
\node[black] at (-0.4,-2.6) {\scriptsize $w$};
\node[black] at (4.6,2.4) {\scriptsize $w$};
\node[black] at (4.4,-2.4) {\scriptsize $w$};
\end{tikzpicture}
\end{tabular}
\captionof{figure}[foo]{On the square lattice displayed on the left, we identify four distinct qubit configurations: $S1$, where all qubits are in the $\lvert-1\rangle$ state; $S2$, with qubits only on the vertical dashed loop in the $\lvert1\rangle$ state, namely $\lvert S2 \rangle = X_v \lvert S1 \rangle$; similarly $\lvert S3 \rangle = X_h \lvert S1 \rangle$; and $\lvert S4 \rangle = X_h X_v\lvert S1 \rangle$. The weights of interest are illustrated on the right.} 
\label{tab:basis in the RBM}
\end{table}

\begin{figure}[ht]
  \centering
  \begin{minipage}[b]{0.48\textwidth}
    \centering
    \includegraphics[width=\textwidth]{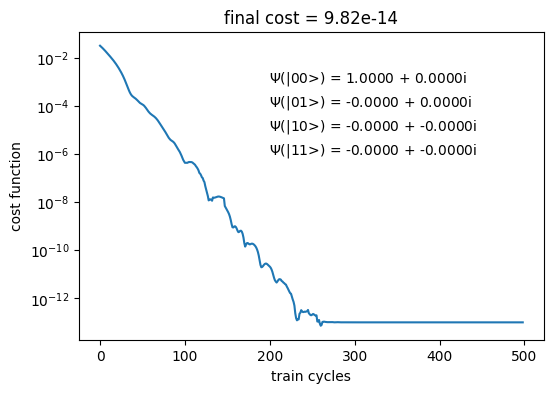}
  \end{minipage}
  \hfill
  \begin{minipage}[b]{0.48\textwidth}
    \centering
    \includegraphics[width=\textwidth]{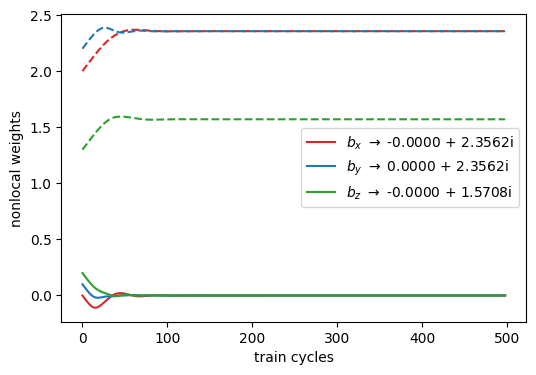}
  \end{minipage}
  \caption{Training results on a $3\times3$ lattice for $\lvert GS\rangle = \lvert00\rangle$.}
  \label{fig:RBM basis 00}
\end{figure}

Let us consider a straightforward example involving the degeneracy state $\lvert 00 \rangle$. We employ the condition $\langle GS \lvert S1 \rangle\!:\!\langle GS \lvert S2 \rangle\!:\!\langle GS \lvert S3 \rangle\!:\!\langle GS \lvert S4 \rangle = 1\!:\!0\!:\!0\!:\!0 $ to analytically determine the weights, yielding in $(b_{x}, b_{y}, b_{z}) = (\frac{3\pi}{4}i, \frac{3\pi}{4}i, \frac{\pi}{2}i)$. Subsequently, we verify the learnability of the RBM, as illustrated in Figure \ref{fig:RBM basis 00}, ensuring that it can accurately and efficiently represent the specified state characteristics.

\begin{figure}[ht]
  \centering
  \begin{minipage}[b]{0.48\textwidth}
    \centering
    \includegraphics[width=1.1\textwidth]{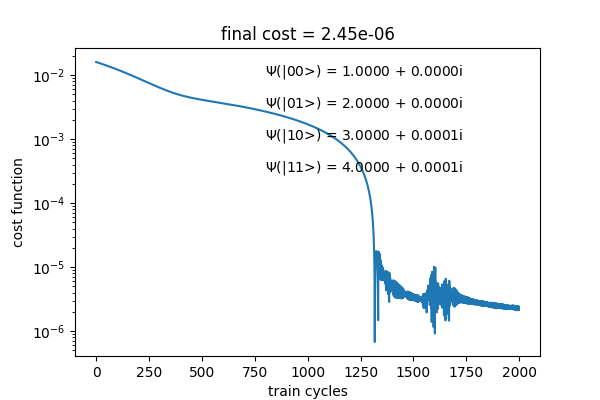}
  \end{minipage}
  \hfill
  \begin{minipage}[b]{0.48\textwidth}
    \centering
    \includegraphics[width=1.1\textwidth]{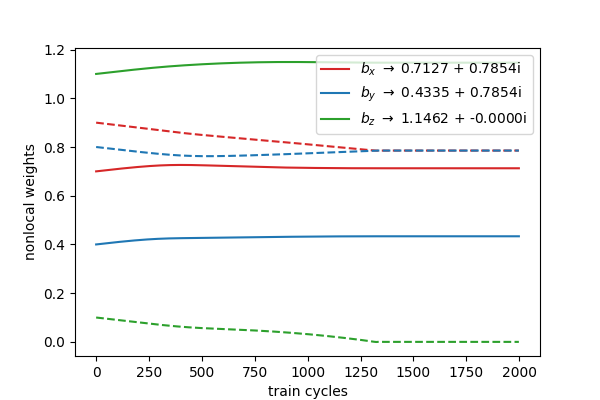}
  \end{minipage}
  \caption{Training results on a $3\times3$ lattice for $\lvert GS\rangle = \lvert00\rangle+2\lvert01\rangle+3\lvert10\rangle+4\lvert11\rangle$.}
  \label{fig:RBM arbitrary 1234}
\end{figure}

\begin{figure}[ht]
  \centering
  \begin{minipage}[b]{0.48\textwidth}
    \centering
    \includegraphics[width=1.1\textwidth]{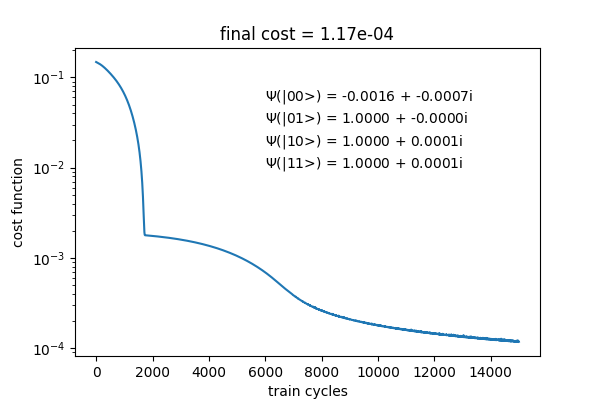}
  \end{minipage}
  \hfill
  \begin{minipage}[b]{0.48\textwidth}
    \centering
    \includegraphics[width=1.1\textwidth]{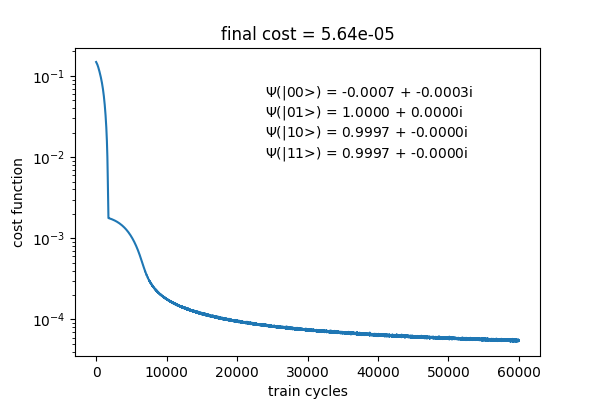}
  \end{minipage}
  \caption{Training results on a $3\times3$ lattice for $\lvert GS\rangle = \lvert01\rangle+\lvert10\rangle+\lvert11\rangle$.}
  \label{fig: approximate zero}
\end{figure}

Similarly, another example with amplitude ratios $\langle GS \lvert 00 \rangle \!:\! \langle GS \lvert 01 \rangle \!:\! \langle GS \lvert 10 \rangle \!:\! \langle GS \lvert 11 \rangle$ $= 1\!:\!2\!:\!3\!:\!4$ results in the solution $(b_{x}, b_{y}, b_{z}) =(\coth^{-1}(2\sqrt{2/3})\!+\!\frac{\pi}{4}i,$ $\coth^{-1}(\sqrt{6})\!+\!\frac{\pi}{4}i, \coth^{-1}(\sqrt{3/2}))$. We then verify the learnability of the RBM, as illustrated in Figure \ref{fig:RBM arbitrary 1234}. We find that finer results can be achieved with smaller training step sizes, though extending training time does not lead to significant improvements. Finally, we present an example that can only be approximated, as shown in Figure \ref{fig: approximate zero}. We observe that finer results are achievable with smaller training step sizes, and unlike the previous case, longer training times also contribute to better outcomes.

\end{document}